\documentclass[%
 reprint,
superscriptaddress,
 amsmath,amssymb,
 aps,prc,
floatfix,
]{revtex4-2}
\usepackage{dcolumn}
\usepackage{bm}

\usepackage[english]{babel}
\usepackage[utf8x]{inputenc}
\usepackage[T1]{fontenc}
\usepackage{lineno}
\usepackage{siunitx}

\usepackage{pstricks}
\usepackage{color}
\usepackage{booktabs}
\usepackage{xspace}

\usepackage[a4paper,top=3cm,bottom=2cm,left=2.5cm,right=2.5cm,marginparwidth=1.75cm]{geometry}

\usepackage{amsmath}
\usepackage{graphicx}

\usepackage[colorinlistoftodos]{todonotes}

\usepackage[colorlinks=true, allcolors=blue]{hyperref}
\def\pp{$pp$\xspace}

\def\pA{$pA$\xspace}

\def\AA{$AA$\xspace}

\newcommand{\smashsim}{S\protect\scalebox{0.8}{MASH}\xspace}

\newcommand{\herwig}{H\protect\scalebox{0.8}{ERWIG} 7\xspace}
\newcommand{\sherpa}{S\protect\scalebox{0.8}{HERPA}\xspace}
\newcommand{\urqmd}{U\protect\scalebox{0.8}{rQMD}\xspace}
\newcommand{\pythia}{P\protect\scalebox{0.8}{YTHIA} 8.2\xspace}
\newcommand{\pytn}{P\protect\scalebox{0.8}{YTHIA}\xspace}
\newcommand{\pytang}{P\protect\scalebox{0.8}{YTHIA}/Angantyr\xspace}
\newcommand{\pytangur}{P\protect\scalebox{0.8}{YTHIA}/Angantyr+U\protect\scalebox{0.8}{rQMD}\xspace}
\newcommand{\figref}[1]{Fig.~\ref{#1}}
\renewcommand{\eqref}[1]{Eq.~(\ref{#1})}

\def\AA{$AA$\xspace}

\def\pp{$pp$\xspace}
\def\pA{$pA$\xspace}

\def\AA{$AA$\xspace}

\newcommand{\pt}           {\ensuremath{p_{\rm T}}\xspace}

\usepackage{siunitx}
\DeclareSIUnit\clight{\text{\ensuremath{c}}}
\DeclareSIUnit\micron{\micro\metre}
\DeclareSIUnit\mrad{\milli\rad}
\DeclareSIUnit\gauss{G}
\DeclareSIUnit\eVperc{\eV\per\clight}
\DeclareSIUnit\nanobarn{\nano\barn}
\DeclareSIUnit\picobarn{\pico\barn}
\DeclareSIUnit\femtobarn{\femto\barn}
\DeclareSIUnit\attobarn{\atto\barn}
\DeclareSIUnit\zeptobarn{\zepto\barn}
\DeclareSIUnit\yoctobarn{\yocto\barn}
\DeclareSIUnit\nb{\nano\barn}
\DeclareSIUnit\pb{\pico\barn}
\DeclareSIUnit\fb{\femto\barn}
\DeclareSIUnit\ab{\atto\barn}
\DeclareSIUnit\zb{\zepto\barn}
\DeclareSIUnit\yb{\yocto\barn}

\newcommand{\nineH}        {$\sqrt{s}~=~0.9$~Te\kern-.1emV\xspace}
\newcommand{\seven}        {$\sqrt{s}~=~7$~Te\kern-.1emV\xspace}
\newcommand{\twoH}         {$\sqrt{s}~=~0.2$~Te\kern-.1emV\xspace}
\newcommand{\twosevensix}  {$\sqrt{s}~=~2.76$~Te\kern-.1emV\xspace}
\newcommand{\five}         {$\sqrt{s}~=~5.02$~Te\kern-.1emV\xspace}
\newcommand{\twosevensixnn}{$\sqrt{s_{\mathrm{NN}}}~=~2.76$~Te\kern-.1emV\xspace}
\newcommand{\fivenn}       {$\sqrt{s_{\mathrm{NN}}}~=~5.02$~Te\kern-.1emV\xspace}

\newcommand{\GeVc}         {Ge\kern-.1emV/$c$\xspace}
\newcommand{\MeVc}         {Me\kern-.1emV/$c$\xspace}
\newcommand{\TeV}          {Te\kern-.1emV\xspace}
\newcommand{\GeV}          {Ge\kern-.1emV\xspace}
\newcommand{\MeV}          {Me\kern-.1emV\xspace}
\newcommand{\GeVmass}      {Ge\kern-.2emV/$c^2$\xspace}
\newcommand{\MeVmass}      {Me\kern-.2emV/$c^2$\xspace}

\begin{document}
\title{Studies of heavy-ion collisions using PYTHIA Angantyr and UrQMD} 
\author{André Vieira da Silva}
\author{Willian Matioli Serenone}
\author{David Dobrigkeit Chinellato}
\author{Jun Takahashi}
\affiliation{Universidade Estadual de Campinas, S\~{a}o Paulo, Brazil}
\author{Christian Bierlich}
\altaffiliation{Also at: Niels Bohr Institute, University of Copenhagen, Denmark}
\affiliation{Dept. of Astronomy and Theoretical Physics, Lund University, Sweden}

\date{}

\begin{abstract}

In this paper, we show predictions from a new QGP-free, no-equilibration, improved baseline model for heavy-ion collisions. It is comprised of the 
\pytang event generator coupled to \urqmd, as a hadronic cascade simulator, and compared to ALICE and CMS
data from Pb-Pb collisions at $\sqrt{s_{\rm{NN}}}$ = 2.76 TeV. This coupling is made 
possible due to a new implementation of the hadron vertex model in \pytang. 
Hadronic rescattering in \urqmd is shown to lead to a significant suppression of 
mid- to high-$p_\perp$ particle yields that is qualitatively consistent with measurements
of nuclear modification factors. 
We further study the effect of
hadronic rescatterings on high-$p_\perp$ particles by using two-particle correlations and show that
some suppression of away-side jets occurs in the hadronic
phase even without any partonic energy loss. 
Finally, the decorrelation of dijet structures at high momenta also leads to a reduction of the 
elliptic flow coefficient $v_{2}\{2\}$. These findings suggest that significant jet-quenching-like effects
may still originate in the hadronic, as opposed to the partonic phase and prove that the usual Pb-Pb baseline, composed of
a superposition of incoherent pp collisions, ignores coherent phenomena that are not strictly related to the QGP 
but may still be highly relevant.

\end{abstract}
\keywords{Alpha, Beta, Delta}

\maketitle

The ultra-relativistic heavy ion (\AA) collisions, measured at the LHC and RHIC experiments,
produce the hottest, densest state of matter available in a laboratory. Such collisions
are expected to lead to a deconfined state of quarks and gluons, denoted the 
Quark--Gluon Plasma (QGP) \cite{Adcox:2004mh,Adams:2005dq,Back:2004je,Arsene:2004fa}. 
Several signatures of this phase of matter have been found in \AA collisions. Notably, 
the yields of high-$p_\perp$ particles are suppressed with respect to expectations from scaled
proton--proton (\pp) collisions \cite{Acharya_2018_RAApaper}, away-side jets are suppressed in central \AA collisions \cite{Adler_2003, PhysRevLett.91.072304} and
particles are emitted anisotropically in azimuth because of collective flow developed during the system evolution \cite{PhysRevC.87.014902}. 

Recently, effects normally associated with the formation of a QGP phase, such as multi-particle flow \cite{Khachatryan:2016txc} and enhanced strangeness
production \cite{ALICE:2017jyt}, were also observed in \pp collisions.
The discovery that the demarcation between QGP-producing and no-QGP-producing collision systems is not as
clear as previously thought, should naturally lead to questions 
regarding the no-QGP baseline used in jet quenching searches in \AA collisions. Not only is it unclear that a given \pp collision can be taken as a pure baseline result, it is
also unclear that an \AA collision can be taken as a simple superposition of \pp collisions, even
without considering QGP effects. It has recently been shown by the ALICE experiment \cite{Acharya:2018hhy} that
the charged-particle multiplicity in very central \AA collisions breaks the scaling with number
of participating nucleons. Furthermore, a correct description of basic
quantities must take into account effects not arising from the QGP phase, but rather from 
nuclear shadowing or diffractive contributions. Additionally, it is well
known that, compared to \pp collisions, the large geometry of an \AA collision also potentially 
allows hadronic rescattering effects to affect relative production rates and kinematics \cite{nucl-th/9803035}, another effect in \AA collisions not linked to QGP production. In this work, we present an improved, QGP-free
baseline for heavy-ion collisions to replace the traditional incoherent superposition of pp collisions. 
This effort is crucially important to disentangle effects that are exclusively due to the QGP.
Special emphasis will be given to the mid- to high $p_\perp$ region  $p_\perp > 4$ GeV/$c$.
 
 In the traditional heavy-ion picture, high-$p_{\perp}$ particles will be 
sensitive to partonic energy loss while traversing the QGP and a modification of the high-$p_{\perp}$
yield in \AA collisions compared to \pp collisions is generally taken as a sign of a QGP phase. This is
quantified by the nuclear modification factor:
\begin{equation}
	\label{eq:raa}
	R_{AA} = \frac{\left. \mathrm{d}^{2}N_{\mathrm{ch}}/\mathrm{d} p_\perp \mathrm{d} y \right|_{AA}}
	{N_{\mathrm{coll}}\left. \mathrm{d}^{2}N_{\mathrm{ch}}/\mathrm{d}p_\perp \mathrm{d} y \right|_{pp} },
\end{equation}
where $y$ is the rapidity and $N_{\mathrm{coll}}$ is a scaling factor corresponding to the number of
binary nucleon--nucleon collisions calculated using the Glauber model \cite{Glauber:1955qq,Miller:2007ri}.

The standard picture of high-$p_\perp$ modification can be phrased in terms of the QGP transport coefficient, which denotes the broadening of the transverse momentum distribution per unit length $\hat{q} = \langle p^2_\perp \rangle_L/L$. According to a selection of the models implemented in the JETSCAPE generator framework \cite{Cao:2017zih}, a high virtuality $Q^2 \gg \sqrt{\hat{q}E}$ (where $E$ is the parton energy) parton will undergo a 
medium-modified DGLAP evolution \cite{Majumder:2009zu} and at lower $Q^2$ the shower modifications can be calculated by transport theory \cite{Schenke:2009gb,He:2015pra}. The JEWEL approach \cite{Zapp:2013vla} treats high to intermediate virtuality by a combination of partonic rescattering and the Landau-Pomeranchuk-Migdal effect \cite{Zapp:2012ak}. In both cases the relevant degrees of freedom are partonic.

\begin{figure}[b]
\begin{center}
	\includegraphics[width=1.0\columnwidth]{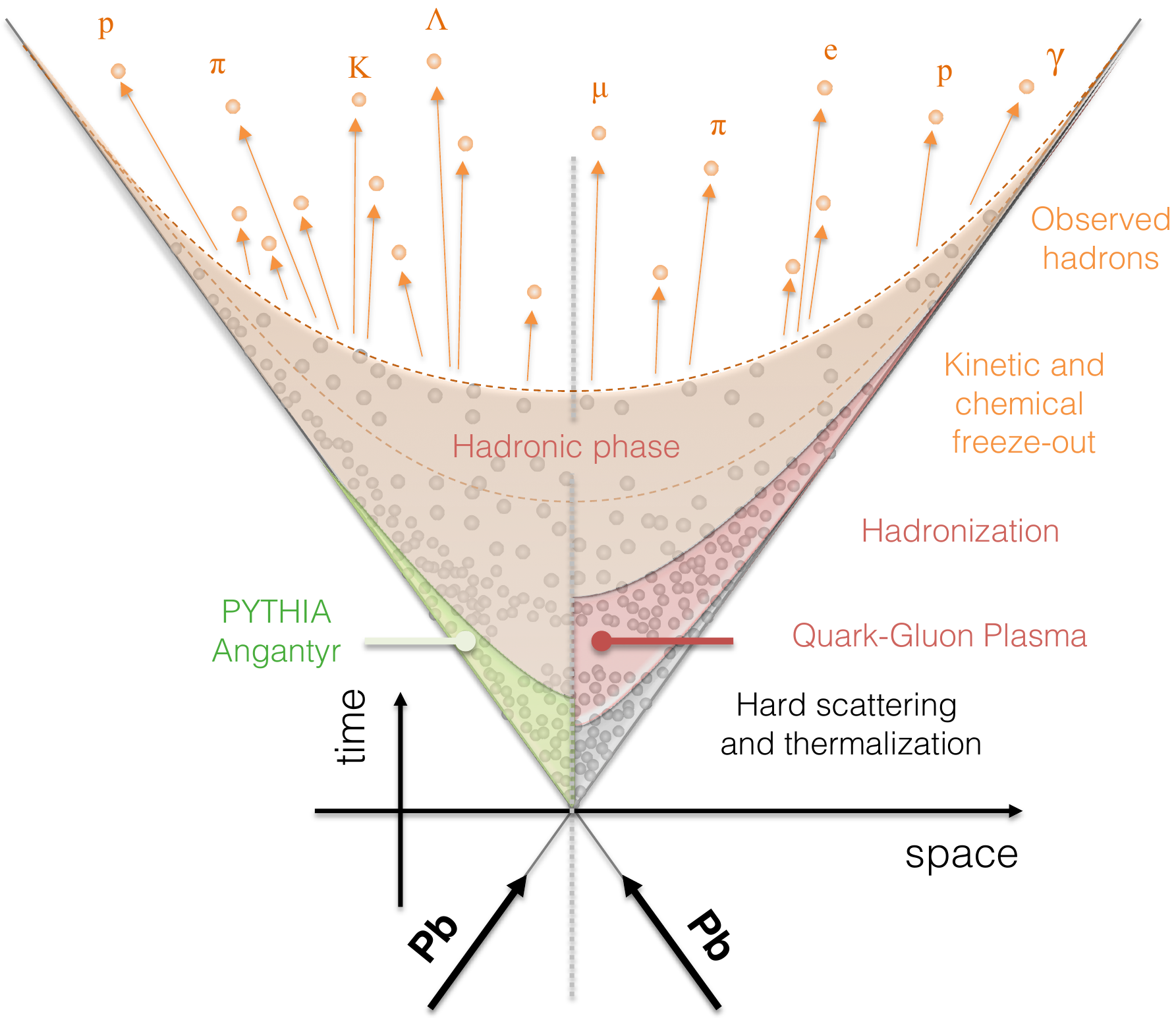}
\end{center}
	\caption{\label{fig:plasma-free}Schematic representation of the modeling of a heavy-ion collision using the usual approach (right side of the figure) and the \pytangur no-QGP baseline method (left side of the figure).}
\end{figure}

Dynamical models based on perturbative QCD coupled to either string or cluster
hadronization \cite{Andersson:1983ia,Marchesini:1983bm} have worked very well to describe most features of $e^+e^-$, $ep$ and \pp collisions.
Such models are implemented in so-called General Purpose Monte Carlo event
generators, of which \pythia \cite{Sjostrand:2014zea}, \herwig \cite{Bellm:2015jjp} or \sherpa \cite{Gleisberg:2008ta} are prominent examples. The \pytn model 
for multi-parton interactions (MPIs) \cite{Sjostrand:1987su} has recently been extended to heavy ion collisions, and the 
resulting \pytang \cite{Bierlich:2016smv,Bierlich:2018xfw} is a QGP-free simulation 
of heavy ion collisions that includes the contributions mentioned above.
For the final state hadronic interactions, quantum
molecular dynamics models such as \urqmd \cite{nucl-th/9803035,hep-ph/9909407} 
and more recently \smashsim \cite{Weil:2016zrk}
have worked very well in hybrid approaches together 
with hydrodynamics for modelling QGP created in heavy ion collisions \cite{0806.1695}, in air-shower simulations \cite{Heck:1998vt} 
as well as detector simulations \cite{Agostinelli:2002hh}. In this paper we present a hybrid approach 
linking \pytangur to model a realistic, QGP free final state of a heavy ion collision. 

\section{Model setup}
The \pytangur hybrid model is outlined in \figref{fig:plasma-free}. On the right side of the figure, the 
standard view on a heavy ion collision is sketched. Importantly, this includes the formation of a strongly coupled, 
nearly thermal QGP phase. This phase can be described by relativistic hydrodynamics \cite{Weller:2017tsr},
and is observed to ``quench'' jets by comparing the single particle yields at high $p_{\perp}$ to that of \pp collisions.
The alternative, QGP free, description is sketched on the left side of \figref{fig:plasma-free}. Instead of assuming the
formation of a QGP, that part of the evolution is carried out using just the \pytang model, which will be presented below. 
The right and left sides of \figref{fig:plasma-free} share the final phase of the evolution, consisting of hadronic 
interactions.

\subsection*{The \pytn model for MPIs}
In a \pp collision, MPIs are generated under the assumption that different partonic interactions are almost 
independent~\footnote{MPIs are not fully independent, as 1) the parton density function corresponding to an 
extracted parton is rescaled by a factor $(1-x)$, and 2) the full collision has to conserve energy and momentum.}.
As such, MPIs are selected according to the $2\rightarrow 2$ perturbative parton--parton cross section:
\begin{equation}
\label{eq:mpi-model}
  \frac{d\sigma_{2\rightarrow 2}}{dp_{\perp}^2} \propto
  \frac{\alpha_s^2(p_{\perp}^2)}{p_{\perp}^4} \rightarrow
  \frac{\alpha_s^2(p_{\perp}^2 + p_{\perp 0}^2)}{(p_{\perp}^2 + p_{\perp 0}^2)^2}.
\end{equation}
Since the cross section diverges like $1/p^4_\perp$, it is regularised using a parameter $p_{\perp 0}$. This value can be interpreted as being
proportional to the inverse of a maximal (colour) screening length of a proton. MPIs are 
interleaved \cite{Corke:2010yf} with the initial state and final state parton showers (ISR and FSR, where `R' stands for radiation), which are also 
ordered in $p_{\perp}$. This implies that MPIs, ISR and FSR obey a combined evolution equation that determines the $p_{\perp}$ of the next step in the evolution, whether that is the generation of a new MPI, 
and ISR or an FSR emission.

The final step before hadronization concerns color reconnection of partonic systems. The idea originates in the notion
that having many MPIs leads to many color strings, which moving from the $N_c \rightarrow \infty$ limit to $N_c = 3$, can be 
connected in many different ways. Since nature is expected to favour configurations with the lowest potential energy -- 
and thus the smallest total string length -- this is the guiding principle for present models. The current default model
for color reconnection, the one used in this paper, gives each MPI system a probability to reconnect with a harder
system which is:
\begin{equation}
	\label{eq:colorrec}
	\mathcal{P} = \frac{p^2_{\perp\mathrm{Rec}}}{p^2_{\perp\mathrm{Rec}} + p^2_\perp},\ \ \ \  
	p_{\perp\mathrm{Rec}} = R \times p_{\perp 0},
\end{equation}
where $R$ is a tunable parameter, and $p_{\perp 0}$ is the same parameter as in \eqref{eq:mpi-model}. This
makes it easier to connect low-$p_{\perp}$ systems with high $p_{\perp}$ ones, essentially making high-$p_{\perp}$
systems ``sweep up'' some of the low-$p_{\perp}$ ones.

\subsection*{\pytang for heavy ion collisions}
In a heavy ion collision, each (projectile) nucleon can interact with several (target) nucleons. 
The amount of interacting nucleons can be determined by a Glauber model, 
to which \pytang makes several additions. Most importantly, for this paper, there is a distinction between different types of 
nucleon--nucleon interactions: elastic, diffractive and absorptive (ie. inelastic, non-diffractive) sub-collisions.
In \pytang this is done by parametrizing the nucleon--nucleon elastic amplitude in impact parameter space ($T(\vec{b})$),
and its fluctuations. Details of the parametrization can be studied in ref. \cite{Bierlich:2018xfw}, but crucially, it allows
for a) determination of all parameters from fits to semi-inclusive \pp cross sections, and b) calculation
of the amplitude $T_{kl}(b)$ for any combination of projectile state $k$ and target state $l$.
Once it is determined which nucleons \textit{may} interact and their type of interaction,
it is then determined if they \textit{will} interact, and, for absorptive sub-collisions, if the interaction
will be considered \textit{primary} or \textit{secondary}. This final distinction is based on the wounded nucleon model \cite{Bialas:1976ed}, according to which a wounded nucleon in a \pA or \AA collision contributes to the final state multiplicity as:
\begin{equation}
	\label{eq:wounded}
	\frac{dN_{\mathrm{ch}}}{d\eta} = w_pF(\eta) + w_tF(-\eta),
\end{equation}
where $w_i$ denotes the number of wounded nucleons in projectile and target respectively, and $F(\eta)$ is a single-nucleon 
emission function (of pseudo-rapidity). For a \pp collision $w_p = w_t = 1$. A proton--deuteron collision, with all nucleons wounded, will
thus reduce to a \pp collision, plus an additional wounded nucleon contributing only on the deuteron side. In the
language of \pytang: one \textit{primary} absorptive collision and one \textit{secondary} absorptive collision. 
The primary collision is modelled as a normal inelastic, non-diffractive collision, whereas the secondary absorptive one 
is treated with inspiration from the Fritiof model \cite{Andersson:1986gw}. In Fritiof, a single string with a mass 
distribution $\propto dM^2/M^2$, similar to diffractive excitation, was used. In \pytang such sub-systems are allowed to 
have MPIs in secondary collisions, following the interleaved MPI/shower prescription described above.
This procedure can be generalized to an arbitrary \AA collision. In \pytang this is done by first ordering all possible 
interaction in increasing local impact parameter. Going from smallest to largest impact parameter, an interaction 
is labelled primary if neither of the participating nucleons have participated in a previous interaction and secondary otherwise.

On the conceptual level, there are some differences between \pp events generated with the \pythia MPI model and
\pA or \AA events generated with the \pytang model. Most importantly, color reconnection is in \pytang only applied 
at the level of individual nucleon--nucleon collisions. As seen from \eqref{eq:colorrec}, the current color reconnection
model includes no dependence on impact parameter, but has instead the implicit assumption that a soft MPI will have a
large spatial spread, and thus be easier to reconnect. While this may be reasonable for a \pp collision where everything
is confined to the transverse size of a single nucleon, it is inappropriate
for heavy ion collisions. In that case, possible reconnections
across separate nucleon--nucleon interactions are therefore neglected altogether. Other more recent developments in string models, such as color ropes \cite{Bierlich:2014xba} and string shoving \cite{Bierlich:2017vhg}, are also not considered in this work. While both effects would be interesting to study with the scope of constructing a full, microscopic alternative to the QGP, the goal here is rather to demonstrate the behaviour of heavy ion collisions considering no parton level QGP-like interactions.
Furthermore the concept of interleaved evolution is also only followed for individual nucleon--nucleon
collisions, and not the full \AA collision.

\subsection*{Hadron production vertices}
After construction of a parton-level event, as outlined above, the color reconnected strings hadronize. This
is done using the Lund string hadronization model \cite{Andersson:1983ia,Andersson:1983jt,Andersson:1998tv,Sjostrand:1984ic}, 
as implemented in \pythia. A string represents the gluonic flux tube stretched between a quark--anti-quark pair. At 
distances larger than $\approx 1$ fm, the confinement potential is linear $V(r) = \kappa r$, with the string 
tension $\kappa \approx 1$ GeV/fm, cf. lattice calculations \cite{Bali:2000un}. As the string grows, energy is transferred
from endpoint momenta to potential energy until the string breaks into hadrons \footnote{For more details about 
fragmentation dynamics, the reader is referred to refs. \cite{Andersson:1983ia,Andersson:1983jt,Andersson:1998tv,Sjostrand:1984ic}, as well as ref. \cite{Sjostrand:2014zea} for a more recent review.}.
A crucial feature of the Lund model is that the string hadronization time, in the string rest frame, can be calculated.
This time signifies directly the end of the pre-hadronic phase, and start of the hadronic cascade. The hadronization time
follows a Gamma distribution with the average:
\begin{equation}
	\langle \tau^2 \rangle = \frac{1 + a}{b\kappa^2}.
\end{equation}
The two parameters $a$ and $b$ also enters the Lund symmetric fragmentation function, which determines the longitudinal 
momentum fraction taken away from the string by each hadron. As such, they are strongly correlated with total charged 
multiplicity and momentum fraction, and can be fitted to $e^+e^-$ collider data. The standard values \cite{Skands:2014pea} gives $\langle \tau^2\rangle \approx 2$ fm/$c$.
In order to couple the output from \pytang to \urqmd, space--time information of hadron production vertices is necessary. 
It was recently shown \cite{Ferreres-Sole:2018vgo} that production vertices of even complicated multi-parton systems can be
extracted in the Lund model and implemented in \pythia. The key component of this translation comes from noting that 
the linear confinement potential gives rise to a linear relationship between space-time and momentum quantities. The
space-time location of a string breakup vertex on a simple $q\bar{q}$ string can therefore be defined as:
\begin{equation}
	\label{eq:spacetime}
	v = \frac{x^+p^+ + x^-p^-}{\kappa},
\end{equation}
where $p^\pm$ are the $q$ and $\bar{q}$ four-momenta, and $x^\pm$ are (normalized) light-cone coordinates of the
break-up point. The production vertex of the hadron is then taken to be the average of the two break-up vertices producing it.
It should be stressed that this simple explanation of a $q\bar{q}$ system does not give justice to the many complications
arising from treatment of multi-parton geometries, gluon loops, massive quarks or junction topologies, but the reader is referred
to ref. \cite{Ferreres-Sole:2018vgo} for details about complicated systems.

The application to \pytang is straightforward. In each nucleon--nucleon collision, each MPI parton is assigned a primordial 
vertex randomly from the overlap of two 2D Gaussian distributions (the nucleon mass distribution), and each nucleon assigned
an overall position in the event according to the initial Glauber simulation.

\subsection*{Coupling to \urqmd} 
Immediately after hadronization, 99.8\% of all particles 
are propagated using \urqmd 3.4 and may interact elastically and inelastically or
decay if unstable, while the remaining 0.2\% are treated separately due
to technical limitations. 
Since hadrons containing charm quarks, which correspond to approximately 
0.2\% of all produced particles, are not adequately treated by 
\urqmd, these particles are exceptionally decayed by PYTHIA and 
only their decay products are used in the hadronic cascade simulation. 
Photons and leptons make up 0.01\% of all particles produced by 
\pytang\ and are discarded prior to the simulation of the hadronic phase as
they would not be recognized or not interact in \urqmd. 

\begin{figure}[tb]
\begin{center}
	\includegraphics[width=1.0\columnwidth]{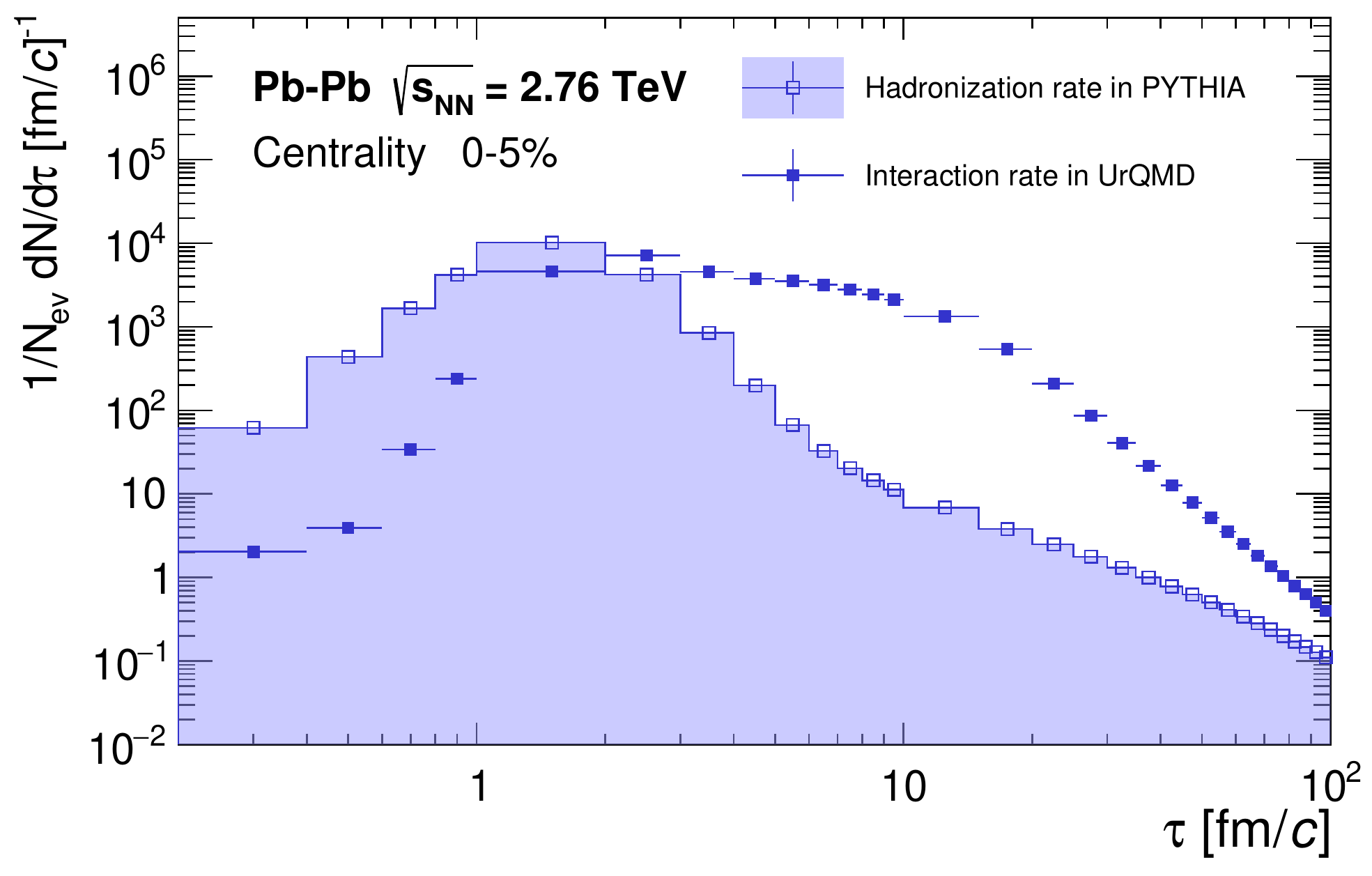}
\end{center}
	\caption{\label{fig:taudistrib}Time $\tau = \sqrt{t^2-z^2}$ distribution 
	of hadrons created by \pytang in 0-5\% central Pb-Pb collisions
	at 2.76~TeV shown together with the time distribution of \urqmd 
	scatterings (solid symbols).}
\end{figure}

\begin{figure}[tb]
\begin{center}
	\includegraphics[width=1.0\columnwidth]{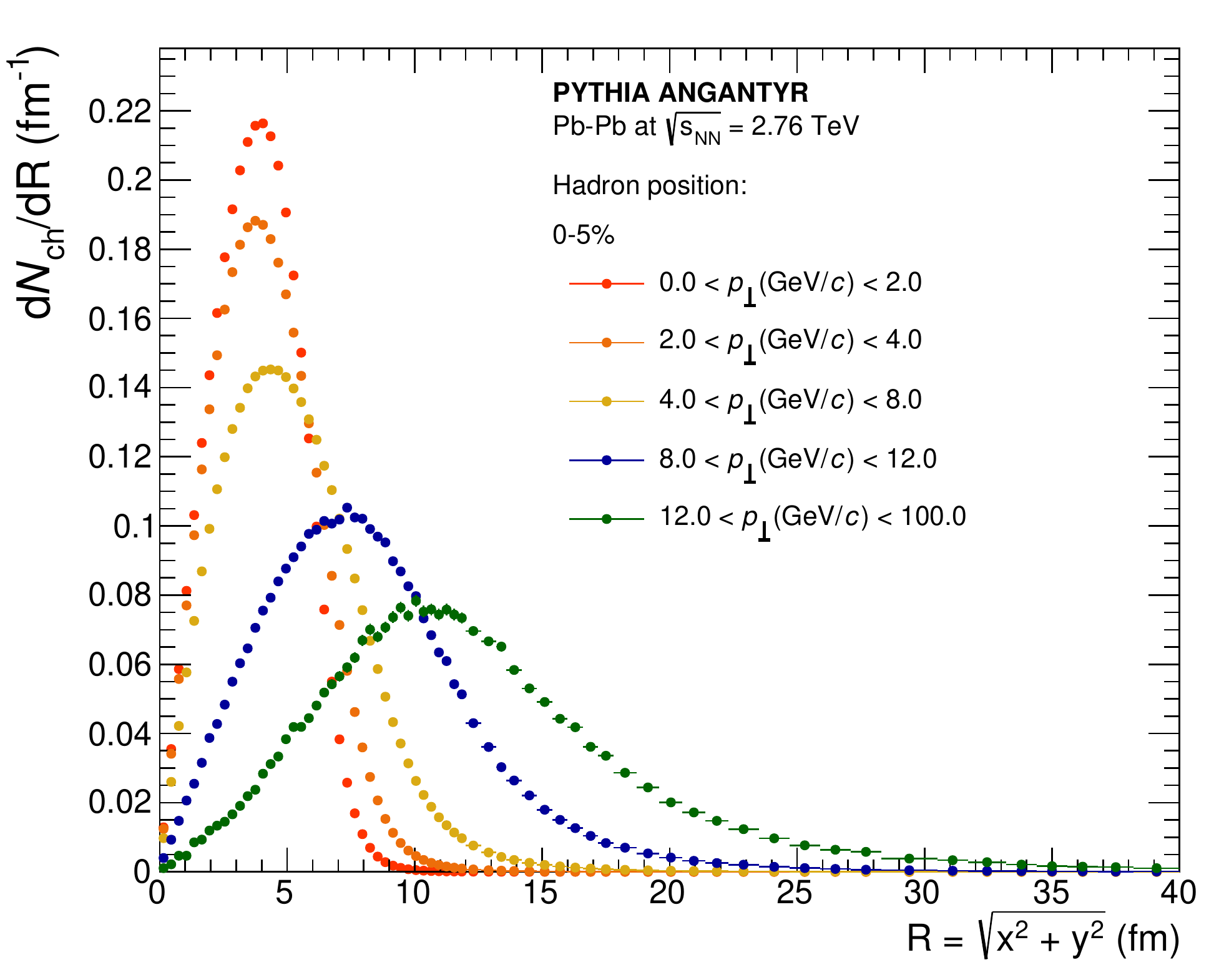}
\end{center}
	\caption{\label{fig:radiusdistrib}Transverse radius $R$ distribution 
	of hadrons created by \pytang in 0-5\% central Pb-Pb collisions
	at 2.76~TeV for various transverse momentum ranges.}
\end{figure}

A notable difference in this coupling, compared to the usual application 
of \urqmd with hydrodynamics-based event generators, is that the latter 
evolve the system by $\mathcal{O}$(10~fm/$c$), while \pytang provides a significant 
fraction of final hadrons already after 1-2~fm/$c$ after the initial hard scatterings 
even in central collisions, as can be seen in Fig.~\ref{fig:taudistrib}. 
Another significant distinction is that, while hydrodynamics-based simulation chains 
produce hadrons at a particlization surface, \pytang provides a position 
distribution over a three-dimensional volume and considers jet-like physical 
correlations between space and time that are not taken into account in hydro-based models. 
Remarkably, in this model, small-$p_{\perp}$ hadrons are all created at low radii in central 
Pb-Pb collisions, high-$p_{\perp}$ 
hadrons will have their production vertices offset from the nucleon--nucleon
collision, as can be seen in Fig.~\ref{fig:radiusdistrib}. 
Ultimately, all these effects lead to an average hadron 
density in \pytang that is approximately 2-3 times larger than the one 
observed in hydrodynamic simulations and the resulting hadronic phase 
lasts significantly longer, as can be seen in the hadron interaction time
distribution in Fig.~\ref{fig:taudistrib}. Therefore, the effects of hadronic interactions 
may be more significant in the \pytangur simulation chain compared to hydrodynamics-based models. 

In order to study simulated events as 
a function of collision centrality, a number of different options, such as 
selecting on impact parameter or charged-particle multiplicity 
in different rapidity ranges, were considered. These options are
found to be largely consistent within 0-70\% centrality. 
The analysis is performed with centrality estimated
using charged-particle multiplicities calculated for the rapidity range
$-3.7<\eta<-1.7$ and $2.8<\eta<5.1$, corresponding to the acceptance of 
the V0M detector of the ALICE experiment and therefore matching what
is done in real data analysis. All results presented in what follows
were produced using a sample of $10^7$ Pb-Pb collisions at
$\sqrt{s_{\rm{NN}}}$ = 2.76~TeV generated with the \pytangur simulation 
chain. 

\section{Results}

\begin{figure}[tb]
\begin{center}
	\includegraphics[width=0.99\columnwidth]{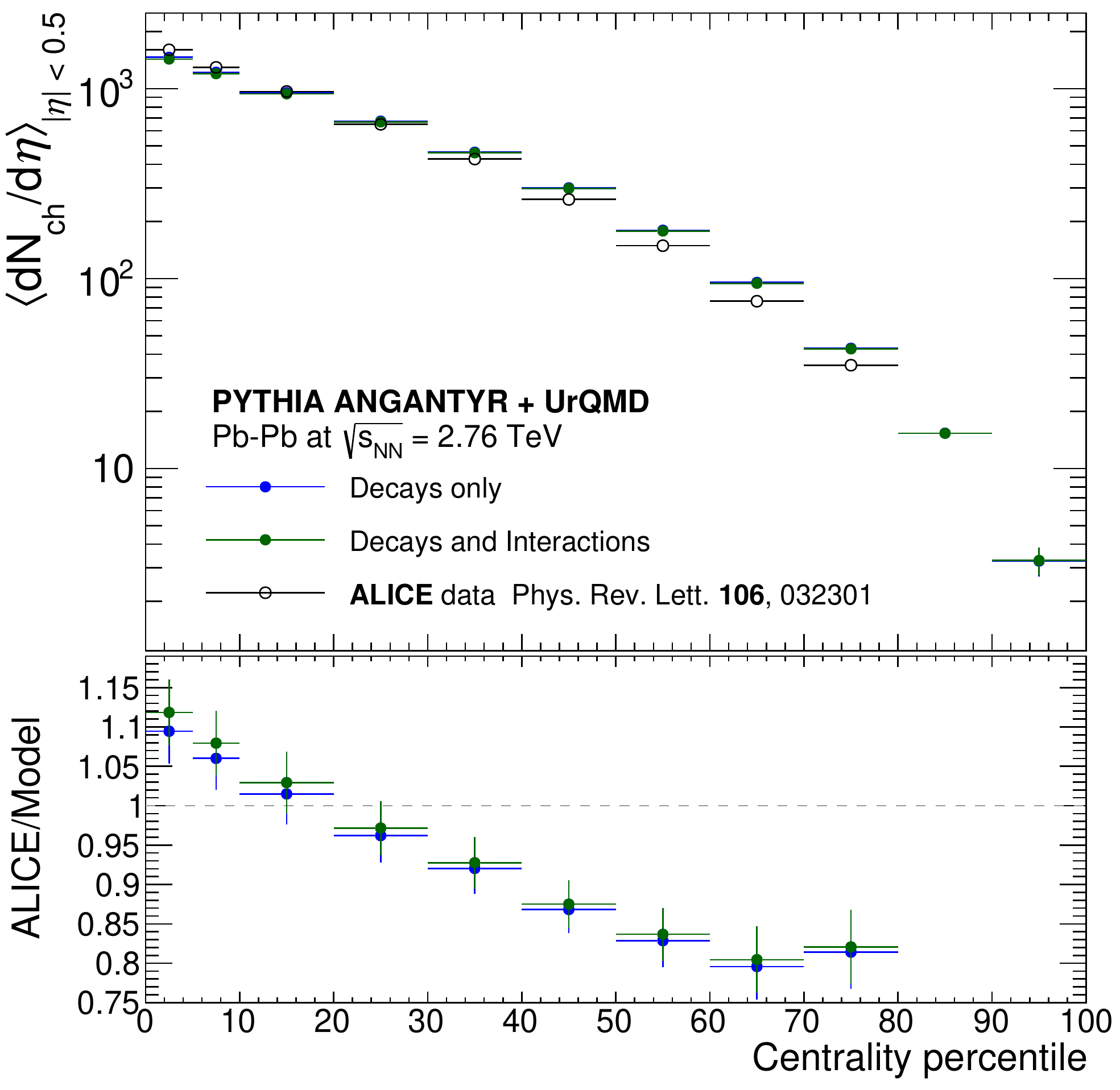}
\end{center}
	\caption{\label{fig:multiplicity}Charged-particle multiplicity density 
	as a function of centrality in Pb-Pb collisions at 2.76~TeV from \pytangur simulations
with and without hadronic scattering compared to ALICE data \cite{1012.1657}.}
\end{figure}

The charged-particle multiplicity density obtained with
the full \pytangur simulation is within approximately 20\% of the corresponding ALICE 
measurements \cite{1012.1657}, as shown in \figref{fig:multiplicity}. 
In order to isolate the effect coming from hadronic scattering, the simulation 
is also re-run with interactions disabled, leading to no more than a 2-3\% 
difference in charged-particle multiplicity densities and therefore no significant change in how 
simulations compare to data. 

\subsection{Particle spectra and $R_{\rm{AA}}$} 

\begin{figure}[tb]
\begin{center}
	\includegraphics[width=0.99\columnwidth]{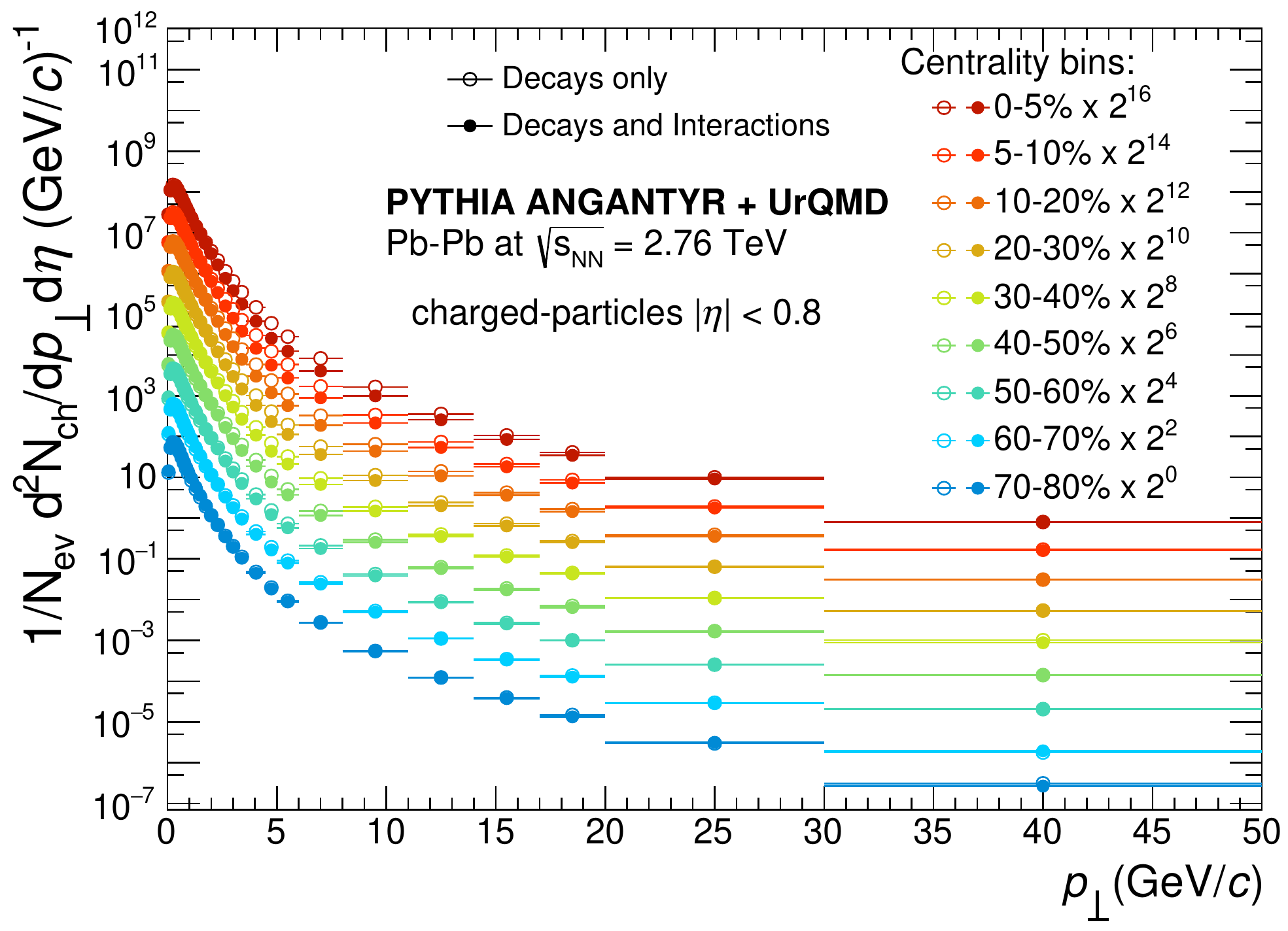}
\end{center}
	\caption{\label{fig:ptspectra}Transverse momentum distributions of charged particles in Pb-Pb 
	collisions at 2.76~TeV simulated with \pytangur.}
\end{figure}

The effect of the hadronic phase can be further characterized by
calculating the $p_{\perp}$ distributions of charged particles, shown 
in \figref{fig:ptspectra} for various centrality classes. 

\begin{figure}[tb]
\begin{center}
	\includegraphics[width=0.99\columnwidth]{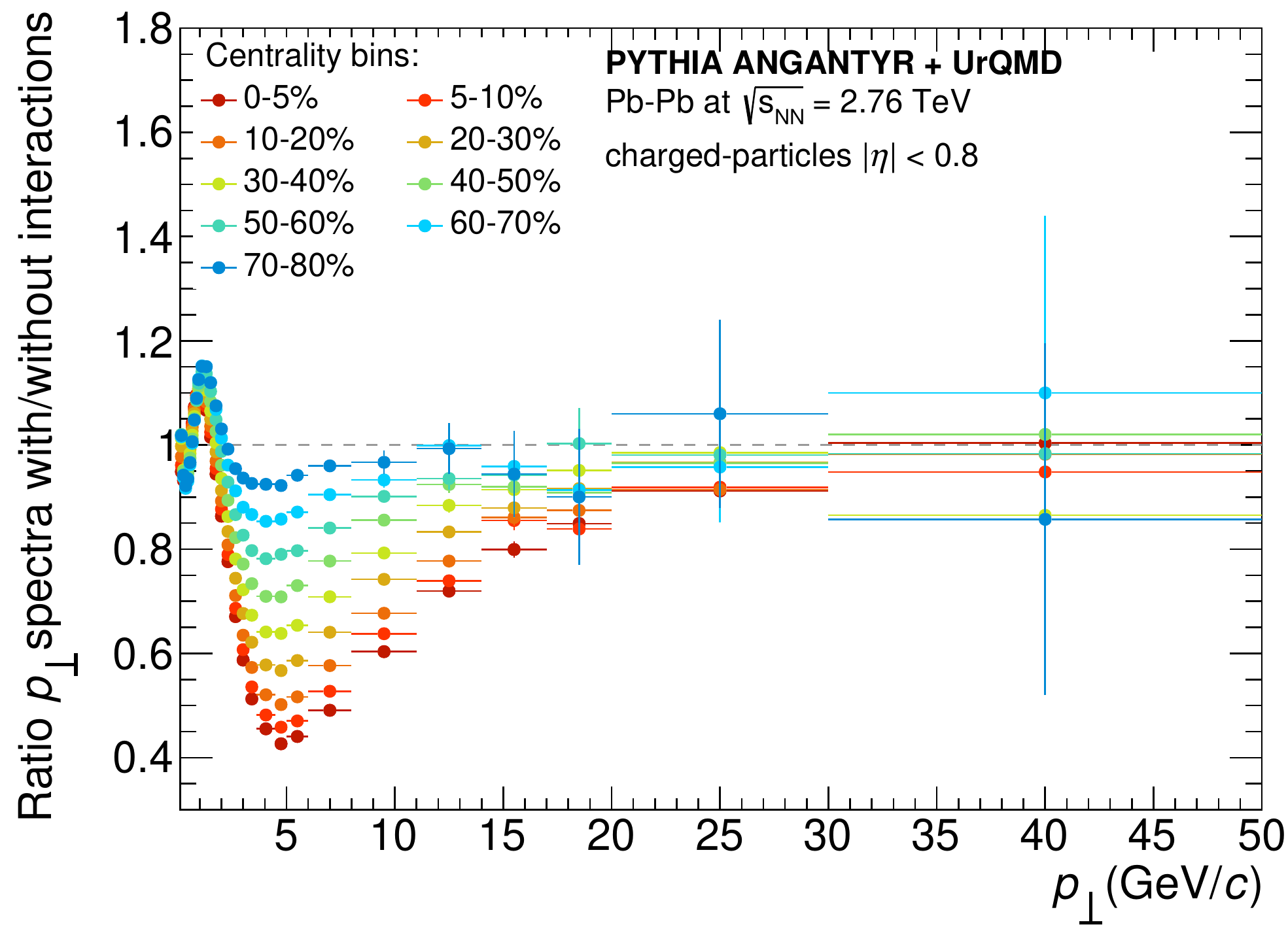}
\end{center}
	\caption{\label{fig:ptspectramod}Ratio of charged particle $p_{\perp}$ distributions with 
	and without hadronic interactions in Pb-Pb collisions at 2.76~TeV simulated with \pytangur 
in various centrality classes.}
\end{figure}

The effect of hadronic interactions can be quantified by calculating the ratio of the $p_{\perp}$ distributions
obtained with and without scatterings, as seen in \figref{fig:ptspectramod}. 
While yields are modified by no more than 10-15\% below 2~GeV/$c$, 
the effect is more pronounced at mid- to high-$p_{\perp}$, with a 
maximum suppression of 60\% taking place at approximately 5~GeV/$c$ 
for 0-10\% collisions. 
The suppression due to hadronic interactions becomes progressively smaller for higher $p_{\perp}$, and
more peripheral collisions. 

It should be noted here that the main model uncertainty related
to coupling \pytang to \urqmd lies in the determination of the vertex
position. The relation in \eqref{eq:spacetime} leads directly to a hadron production point as the average of
two subsequent break-up points. As noted in ref. \cite{Ferreres-Sole:2018vgo}, this definition is not unique, but could differ from the
average up to $\pm p_h/2\kappa$, where $p_h$ is the hadron four-momentum. Taking this at face value, 
the uncertainty of the modification shown in Fig.~\ref{fig:ptspectramod}
is of up to approximately $\pm50\%$ of the fractional values. While this
value is probably an overestimate of the real uncertainty, as very 
extreme definitions were considered, this result also indicates that 
further theoretical work may still be needed to constrain 
hadron creation vertices. 

\begin{figure}[tb]
\begin{center}
	\includegraphics[width=0.45\textwidth]{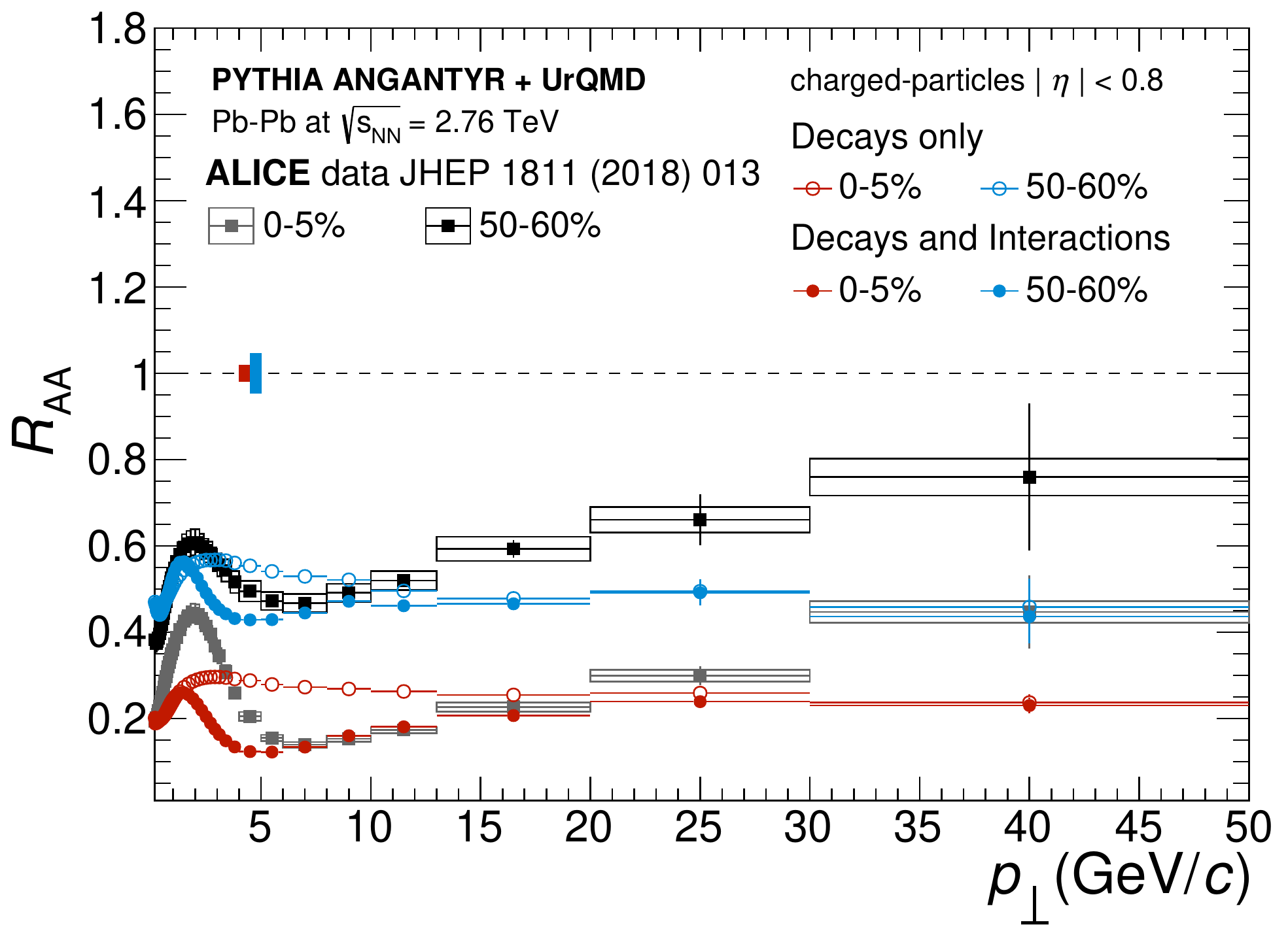}
\end{center}
	\caption{\label{fig:nucmod}Nuclear modification factor ($R_{\rm{AA}}$ in mid-central and central collisions 
	in Pb-Pb at $\sqrt{s_{\rm{NN}}}$ = 2.76~TeV in \pytangur 
	with and without hadronic interactions compared to data
	from the ALICE experiment \cite{1802.09145}.}
\end{figure}

The results seen in \figref{fig:ptspectramod} are reminiscent
of the suppression of high-$p_{\perp}$ particles seen in the nuclear 
modification factor $R_{\rm{AA}}$. To calculate the $R_{\rm{AA}}$ 
from the \pytangur model, pp collisions were generated
using the exact same settings and simulation chain and the average
number of binary collisions $N_{\rm{coll}}$ is assumed to be
the one calculated by the ALICE Collaboration using the Glauber model \cite{ALICE-PUBLIC-2018-011}. 
The simulated $R_{\rm{AA}}$ with 
and without hadronic interactions can be seen in \figref{fig:nucmod} for Pb-Pb 
 at $\sqrt{s_{\rm{NN}}}$ = 2.76 TeV, respectively, 
for two selected centrality classes. 
With interactions disabled, the model overestimates the $R_{\rm{AA}}$ at mid-$p_{\perp}$, from 4 GeV/$c$ to 10 GeV/$c$ for peripheral collisions and 4 GeV/$c$ to 20 GeV/$c$ for the most central collisions. When hadronic interactions are enabled, there is a further suppression of $R_{\rm{AA}}$ in these $p_{\perp}$ regions, improving the agreement of our simulation with the observed experimental data in the $p_{\perp}$ range.
At higher momenta, however, \pytangur fails to describe the data, underpredicting the $R_{\rm{AA}}$ significantly. 
Notably, had we normalized the $R_{\rm{AA}}$ differently, such that the result from \pytang should converge to 
unity at high momenta, then the modification observed when 
enabling \urqmd would lead to a yield suppression that is not strong 
enough to reproduce the measured $R_{\rm{AA}}$. This indicates that there
is no trivial change to the calculation that would lead to a 
correct simultaneous reproduction of the charged-particle multiplicity 
densities and an $R_{\rm{AA}}$ that converges to unity at high $p_\perp$.

This is due to \pytangur being based on the wounded nucleon model, and thus having no concept of $N_{coll}$ scaling
built in. From basic factorization arguments, it is clear that at some scale $p_{\perp,sep}$, $N_{coll}$ scaling is expected.
The numerical value of this separation scale between the hard and the soft component can, however, not be inferred directly.
In the seminal paper by Eskola \textit{et al.} \cite{Eskola:1988yh}, a value of $p_{\perp,\rm{sep}} = 2$ GeV/$c$ was used, while other 
contemporary approaches, such as Fritiof \cite{Andersson:1986gw} or by Ranft \cite{Ranft:1987xn}, would place the value higher,
in line with the results shown in this paper. The significance of the results shown in \figref{fig:nucmod} can thus be interpreted
differently, depending on one's choice of $p_{\perp,s\rm{ep}}$. If $p_{\perp,\rm{sep}}$ is as small as 2 GeV/$c$, the \pytangur baseline is
wrong down to that value. Regardless of the baseline, \figref{fig:nucmod} shows that a hadronic cascade can significantly impact
$R_{\rm{AA}}$ with up to 50\%, but the agreement between simulation and data up to around $p_\perp = 15$ GeV/$c$
is, with this choice of $p_{\perp,\rm{sep}}$, merely co-incidental~\footnote{We note that there is a similar agreement between 
\pytangur and data for $R_{\rm{AA}}$ measured in Xe-Xe collisions at $\sqrt{s_{\rm{NN}}}$ = 5.44~TeV, making a complete co-incidence 
unlikely.}. On the other hand, if $p_{\perp,\rm{sep}}$ is higher, and the hard
component starts taking over at values as high as $p_\perp \gtrapprox 15-20$ GeV/$c$, the result in \figref{fig:nucmod} suggests that no 
QGP modification of high-$p_\perp$ partons is necessary to describe $R_{\rm{AA}}$ up to around that value, and that $R_{\rm{AA}}$ is 
therefore \emph{not} a good observable to study medium modifications in the intermediate $p_\perp$ range.

\subsection{Two-particle correlation study} 

To further understand the nature of the high-$p_{\perp}$ yield suppression, we 
performed two-particle correlation (2PC) studies using a high-$p_{\perp}$ trigger range of 
6 to 8 GeV/$c$ and an associated trigger range of 4 to 6 GeV/$c$, corresponding to the
region of maximum suppression observed in Fig.~\ref{fig:ptspectramod}. The correlation 
function $\rm{C}(\Delta\phi,\Delta\eta)$ is calculated for particles within $|\eta|$ < 2.5 
and is defined as
\begin{equation} 
\label{corrfunc}
\rm{C}_{\rm{full}}(\Delta\phi,\Delta\eta) = \frac{\rm{C}_{\rm{same}}(\Delta\phi,\Delta\eta)}{\alpha\times\rm{C}_{\rm{mixed}}(\Delta\phi,\Delta\eta)},
\end{equation}
where $\rm{C}_{same}$ denotes the correlation function calculated with pairs from the same event
and $\rm{C}_{mixed}$ is calculated for particles from different events, following the `mixed event'
technique. 
This technique corrects for limited pair acceptance, and is often used by experiments. It is used here
in order to repeat the analysis in the same way as an experiment would perform it.
Events are only mixed if they differ in centrality by no more than 2\% and $\rm{C}_{\rm{mixed}}$
is populated using at least 500 unrelated events for each single event being processed. The normalization
factor $\alpha$ is chosen such that $\rm{C}_{\rm{mixed}}(\Delta\phi,\Delta\eta)$ is unity at
$\Delta\eta=0$, indicating that particles having the same $\eta$ value will always be accepted by
the $|\eta|<2.5$ acceptance cut and be paired successfully. 
Furthermore, uncorrelated background in $\rm{C}$ is calculated using
the mixed event technique as well, though for that case events are mixed after rotating them such 
that their event planes (EPs), defined as the plane constructed out of the beam axis and the vector connecting
the two nuclei centers in the transverse plane, are aligned. The background to be subtracted from 
$\rm{C}(\Delta\phi,\Delta\eta)$ is then calculated using
\begin{equation} 
\label{corrfunc2}
\rm{C}_{bg}(\Delta\phi,\Delta\eta) = \frac{\beta\times\rm{C}_{\rm{mixed}}^{\rm{aligned~ EP}}(\Delta\phi,\Delta\eta)}{\alpha\times\rm{C}_{\rm{mixed}}(\Delta\phi,\Delta\eta)},
\end{equation}
where $\alpha$ and $\rm{C}_{mixed}$ are the same as in Eq.~(\ref{corrfunc}), $\rm{C}_{\rm{mixed}}^{\rm{aligned~EP}}$
is the mixed-event correlation function calculated with aligned EPs and $\beta$ is a normalization factor to account
for imperfections in the mixed-event background estimate. The factor $\beta$ is calculated by matching the particle
yield of $\rm{C}_{\rm{bg}}$ to that of $\rm{C}_{\rm{full}}$ in the near-side region 
but away from the peak, sampled in $1.0<|\Delta\eta|<4.0$ and $|\Delta\phi|<\pi/2$. The final, background-subtracted
correlation function is then defined simply as: 
\begin{equation} 
\label{corrfunc3}
\rm{C}(\Delta\phi,\Delta\eta) = \rm{C}_{\rm{full}}(\Delta\phi,\Delta\eta) - \rm{C}_{bg}(\Delta\phi,\Delta\eta)
\end{equation}
This subtraction method is
perfectly suited to remove elliptic-flow-like correlations, which are in fact present in this model as discussed in the next
section, and any imperfections and higher-order corrections have been tested to be negligible in the momentum 
ranges considered in this study. 


\begin{figure}[tb]
\begin{center}
	\includegraphics[width=0.48\textwidth]{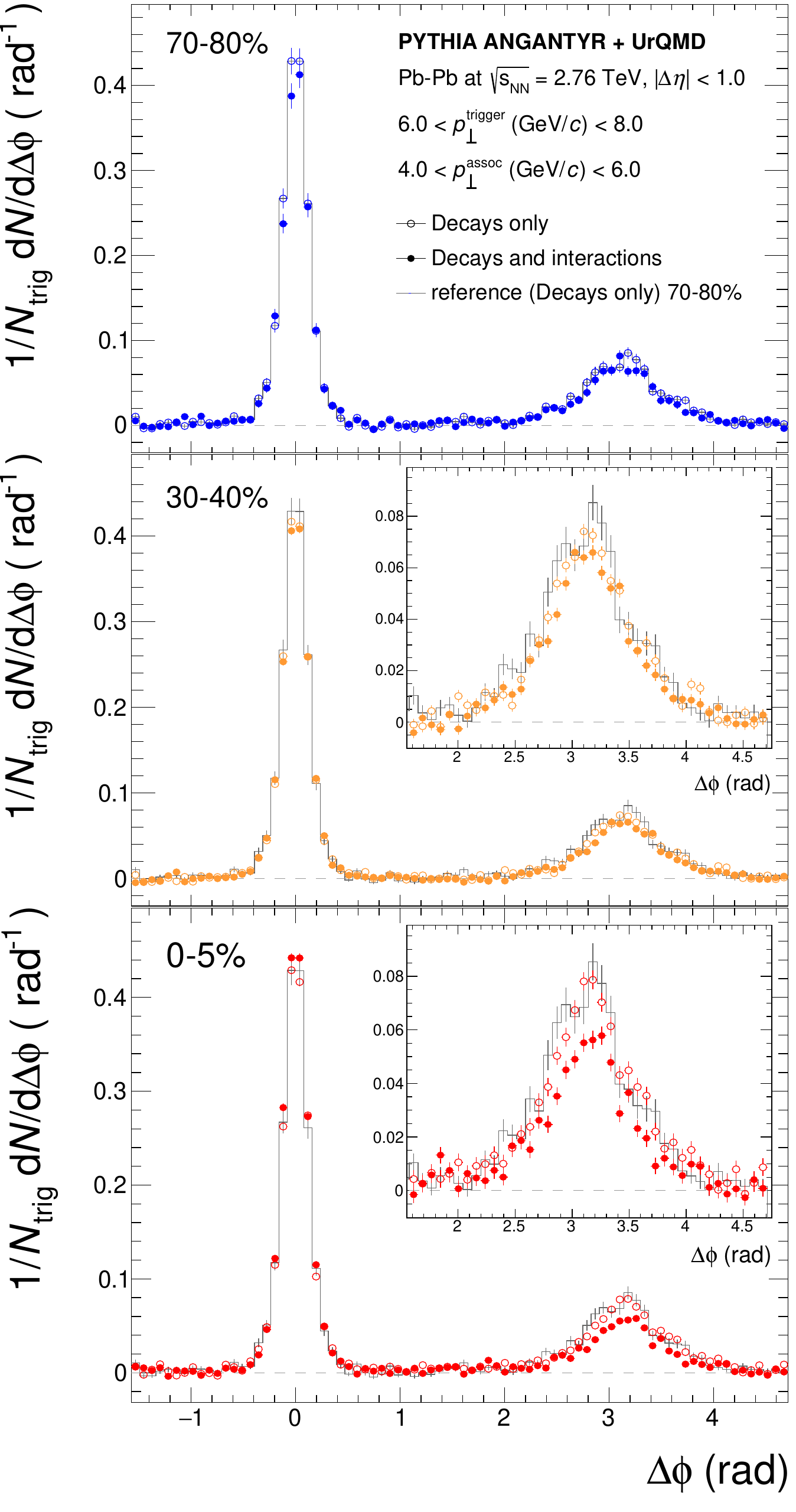}
\end{center}
	\caption{\label{fig:2pc}Two-particle correlation in $\Delta\phi$ using 6.0 < $p_\perp^{\rm trigger}$ (GeV/$c$) < 8.0 
	and 4.0 < $p_\perp^{\rm assoc}$ (GeV/$c$) < 6.0 in three selected event classes
	 in Pb-Pb at $\sqrt{s_{\rm{NN}}}$ = 2.76~TeV in \pytangur 
	with and without hadronic interactions. }
\end{figure}

This procedure was repeated for all centralities analysed before and the results of the 
projection of $\rm{C}(\Delta\phi,\Delta\eta)$ onto $\Delta\phi$ are shown in 
Fig.~\ref{fig:2pc} in a few selected centrality classes. 
A small, but still significant, suppression of the away-side jet structure
can be observed when hadronic interactions are enabled, while the near-side yields are 
only minimally affected even in central events and match those observed in peripheral collisions. 
This is an indication that the hadronic interactions are such 
that an outgoing trigger particle is likely to be closer to the edge of the heavy-ion collision, 
in which case the corresponding away-side jet structure will undergo significant interaction with
the remainder of the system. 

This observation is qualitatively reminiscent of measurements performed at
RHIC by STAR in which dijets were studied using 2PC and the away side was completely suppressed
in central collisions \cite{Adler_2003, PhysRevLett.91.072304}. In that case, the explanation of this phenomenon was also 
related to the outgoing trigger particle being away from the center of the collision, but the 
suppression of the away side was seen as a consequence of partonic rather than hadronic interactions. However, 
it is worth noting that though qualitatively consistent, the overall magnitude of the effect 
observed in our model is smaller than the one required to achieve a full suppression of the away side. 

\begin{figure}[tb]
\begin{center}
	\includegraphics[width=0.48\textwidth]{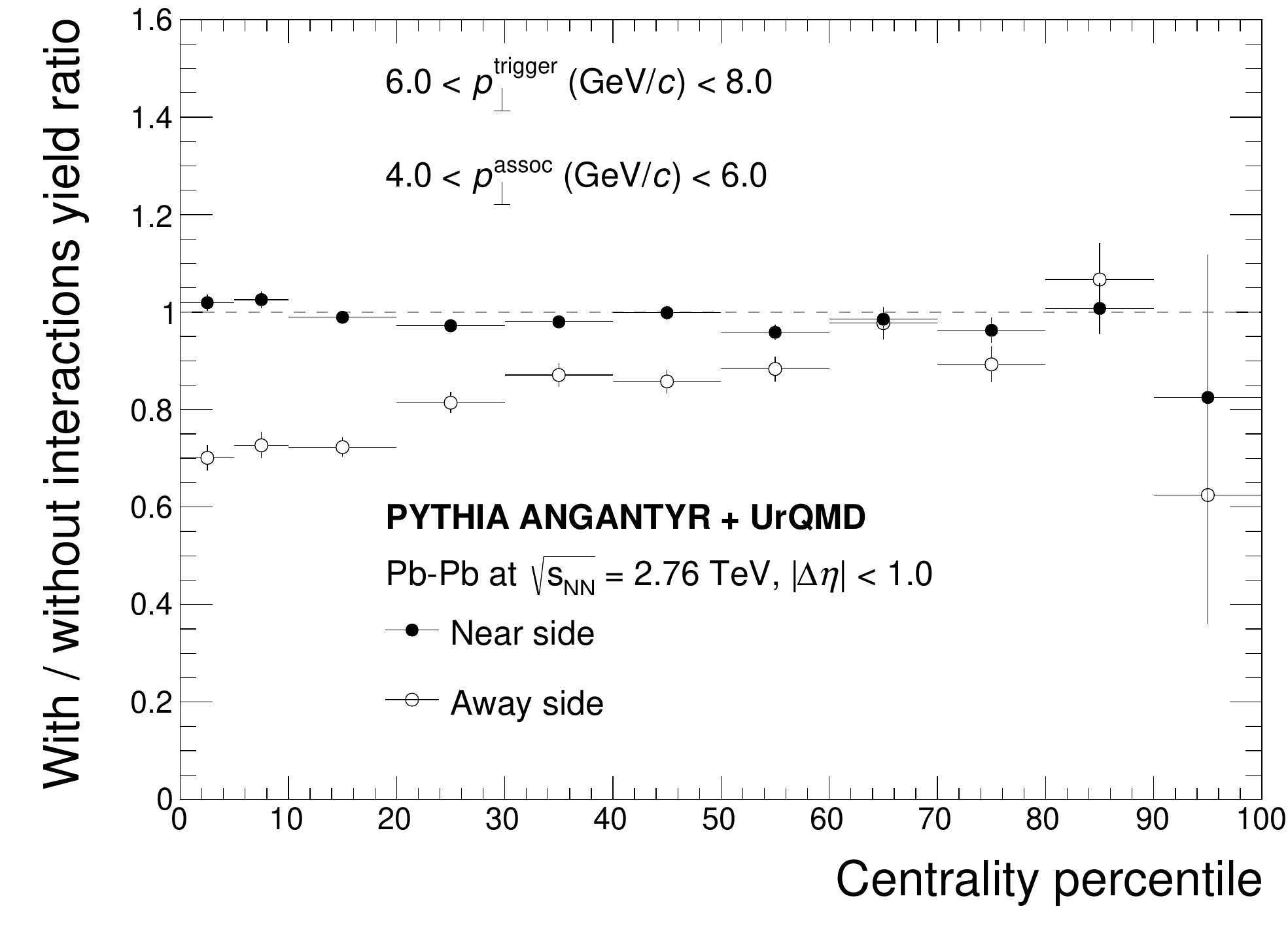}
\end{center}
	\caption{\label{fig:IAA}Jet yield modification for the away- and near-side jets as a function of
	centrality in Pb-Pb at $\sqrt{s_{\rm{NN}}}$ = 2.76~TeV in \pytangur 
	with and without hadronic interactions. }
\end{figure}

To quantify the magnitude of the yield loss in the away side as a function of centrality, 
we integrate the background-subtracted correlation function $\rm{C}(\Delta\phi,\Delta\eta)$ in the
near side, $-\pi/2<\Delta\phi<+\pi/2$, and in the away side, $+\pi/2<\Delta\phi<+3\pi/2$, with 
and without hadronic interactions and then plot the ratio of these two configurations in Fig.~\ref{fig:IAA}. 
While the near-side jet yield is independent of centrality, the away-side yield is progressively
suppressed with centrality, reaching a maximum suppression of approximately 30\% for 0-5\% Pb-Pb 
events. 

\begin{figure}[tb]
\begin{center}
	\includegraphics[width=0.48\textwidth]{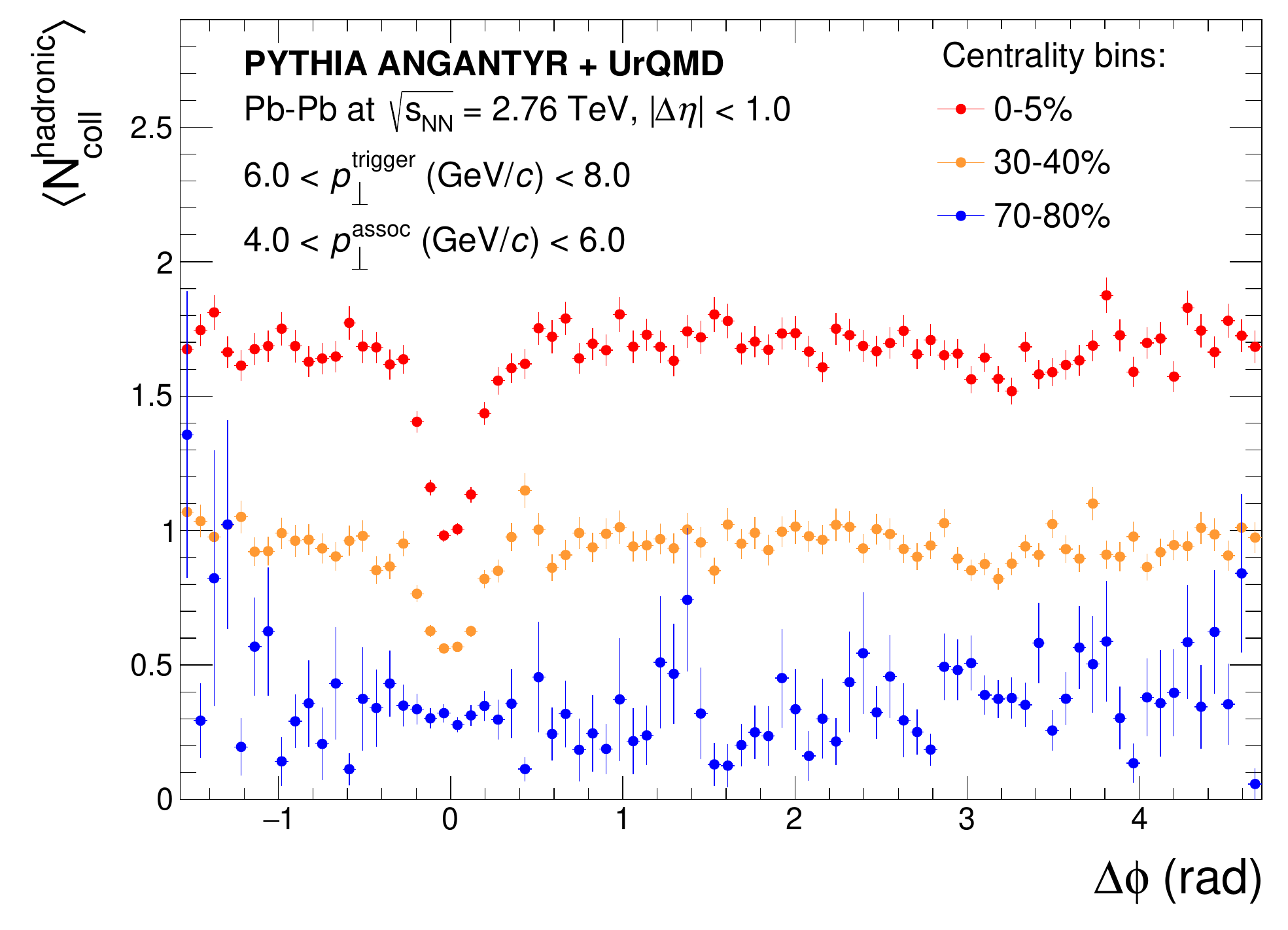}
\end{center}
	\caption{\label{fig:avncoll}Average number of hadronic collisions as a function of 
	$\Delta\phi$ in three selected event class 
	 in Pb-Pb at $\sqrt{s_{\rm{NN}}}$ = 2.76~TeV in \pytangur 
	with and without hadronic interactions. }
\end{figure}

The reason for the observed dynamics can be further explored by calculating the average number of 
hadronic collisions $\langle N_{\rm{coll}}^{\rm{hadronic}}\rangle$ as a function of $\Delta\phi$, as seen 
in Fig. \ref{fig:avncoll}. It is revealed that the $\langle N_{\rm{coll}}^{\rm{hadronic}}\rangle$
is significantly smaller around $\Delta\phi\approx 0$, indicating that the triggered near-side jet
leaves the collision without any interaction. This is not the case in the away-side region $\Delta\phi\approx\pi$, 
suggesting that away-side particles are likely to be lost while traversing the hadronic medium in 
our simulations. 

\subsection{Elliptic flow coefficients}

Another way of characterizing heavy-ion collisions is the study of elliptic flow. In 
traditional hybrid models, elliptic flow arises because of a conversion of geometric anisotropy to a 
momentum space anisotropy that takes place mostly in a partonic phase. It is usually quantified
by the elliptic flow coefficient $v_{2}$, which can be calculated using two-particle cumulants \cite{PhysRevC.63.054906,PhysRevC.64.054901} 
and has been measured extensively by experiments such as ALICE and CMS. When calculated with 
particle pairs in the same rapidity window, the $v_{2}$ obtained with cumulants is denoted $v_{2}\{2\}$ 
and is affected by dijet correlations. 

\begin{figure}[tb]
\begin{center}
	\includegraphics[width=0.47\textwidth]{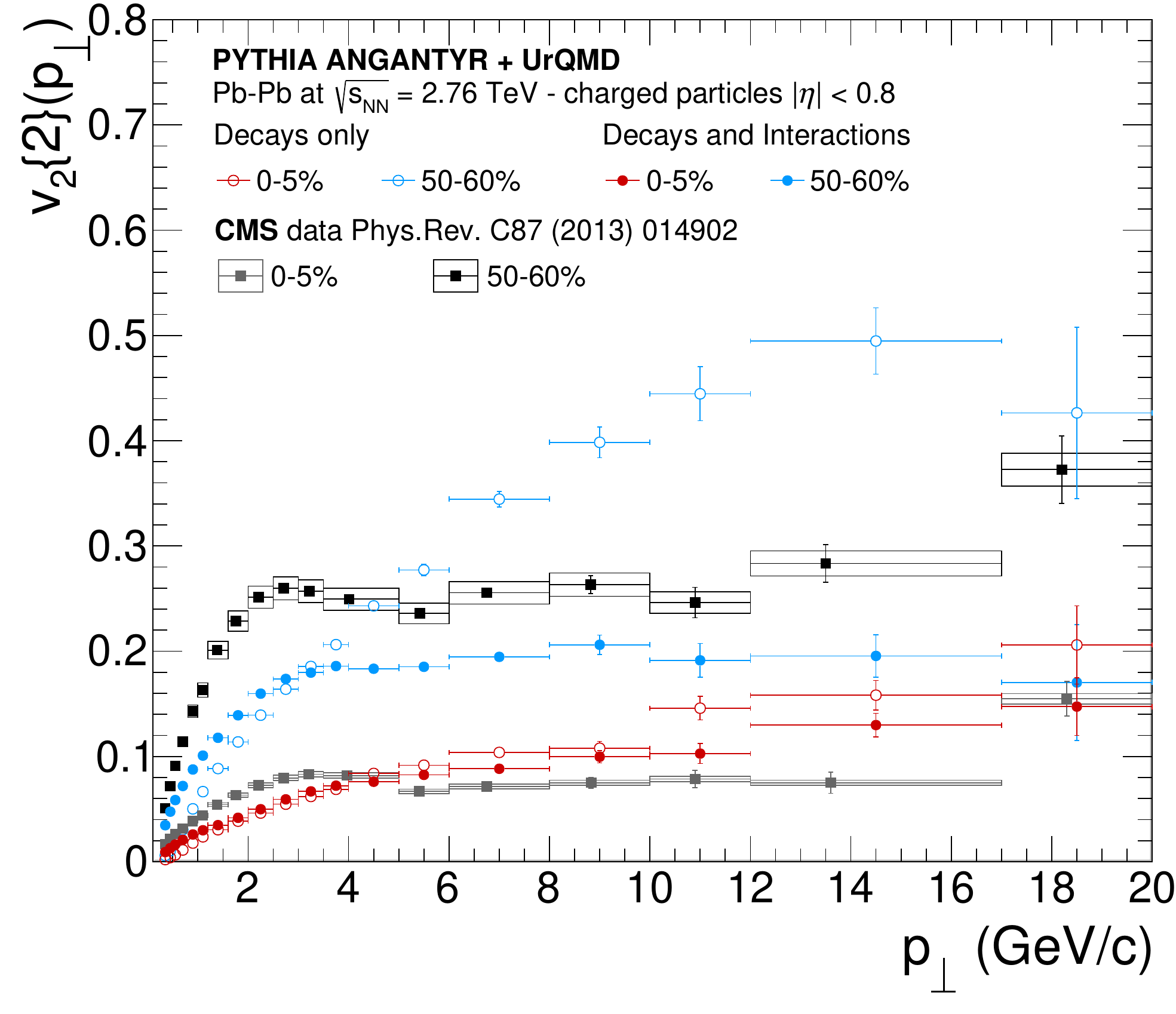}
\end{center}
	\caption{\label{fig:v2}Elliptic flow coefficient $v_{2}$ as a function of $p_{\perp}$ in mid-central and central 
	collisions in Pb-Pb at $\sqrt{s_{\rm{NN}}}$ = 2.76~TeV in \pytangur 
	with and without hadronic interactions compared to data
	from the CMS experiment \cite{PhysRevC.87.014902}.}
\end{figure}

In PYTHIA/Angantyr, particles are produced incoherently and without any correlation to the event plane, hence no elliptic flow is expected. However, non-flow effects such as correlations of decays, momentum conservation and jets will still produce a measurable $v_{2}\{2\}$, as seen in Fig.~\ref{fig:v2}. 
When coupled to \urqmd, the coordinate space anisotropy 
does get converted into a momentum anisotropy via hadronic interactions at low $p_\perp$, even if this effect is not sufficient
to reproduce the measured $v_{2}\{2\}$, as seen in Fig.~\ref{fig:v2}.
This observation is consistent with 
previous work in which hadronic interactions were observed to lead to a momentum anisotropy in the final state \cite{Lu_2006}. 


At high $p_\perp$, where non-flow contributions dominate, the $v_{2}\{2\}$ 
obtained with hadronic interactions is significantly lower than the one observed without interactions. This is because dijet correlations 
are suppressed, as seen in Fig.~\ref{fig:2pc}, such that the behavior observed in \pytangur simulations is more similar to measurements from the
CMS collaboration, at least qualitatively. The reduction of the $v_{2}\{2\}$ is most pronounced in the 50-60\% event class, as shown in Fig.~\ref{fig:v2}, as opposed to in the 0-5\% class. This observation is due to the fact that in central events, non-flow correlations
are so numerous that they are already sufficiently diluted by two-particle combinatorics. 

\begin{figure}[tb]
\begin{center}
	\includegraphics[width=0.47\textwidth]{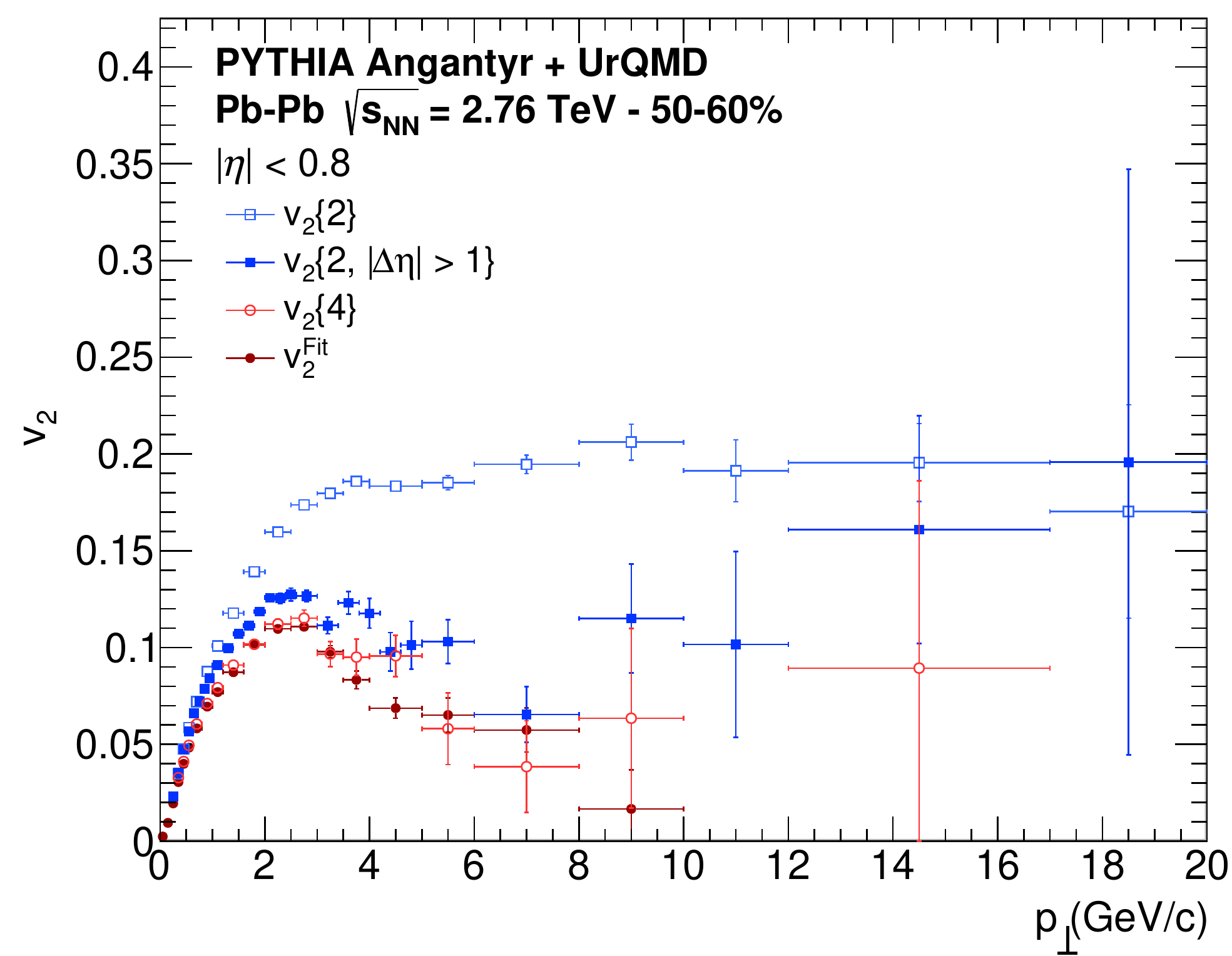}
\end{center}
	\caption{\label{fig:v2_gap} Transverse momentum-differential elliptic flow coefficient $v_{2}$ calculated using various methods in Pb-Pb collisions at \twosevensix. See text for a detailed description.}
\end{figure}

The observation of a non-zero $v_{2}$ in \pytangur with 
hadronic interactions can be further investigated by looking
at further options of calculating the $v_{2}$, as seen in 
Fig.~\ref{fig:v2_gap}. The $v_{2}\{2\}$ calculated with an 
eta gap of 1.0 shows a significant reduction of the flow 
coefficient, especially at high momenta, indicating that 
short-range two-particle correlations dominate in that regime. 
This is further corroborated by the values obtained if 
calculating $v_{2}$ using 4-particle cumulants, i.e. the 
$v_{2}\{4\}$, which is only slightly smaller than the $v_{2}\{2,|\Delta\eta|>1.0\}$, the $v_{2}$ obtained with an eta gap. 
To test if the observed $v_{2}$ is correlated with the event plane, 
we have calculated the $v_{2}$ using an alternate method in which 
the azimuthal particle emission distribution with respect to the
event plane is adjusted with a $A+B\cos{\Delta\phi}$ function, 
where A and B are free parameters and $\phi$ is the angle with 
respect to the event plane. The result is shown as $v_{2}^{\rm{Fit}}$ 
in Fig.~\ref{fig:v2_gap}. The similarity of the $v_{2}^{\rm{Fit}}$ 
with $v_{2}\{4\}$ indicates that indeed the observed elliptic
flow correlates with the event plane direction, as it should also 
in hydrodynamics simulations, despite the fact that the absolute
values are still smaller than data. 

\subsection{Studying UrQMD response\label{lab:toymod}}

Given that the hadronic phase led to a buildup of flow in the PYTHIA-based
model, an important question is if this buildup would behave additively in case
an initial pre-hadronic-phase flow was present. This is relevant for future 
inclusion of collective behavior of the system in the pre-hadronization,
e.g.~string shoving in the \pytang 
model \cite{Bierlich:2016vgw,Bierlich:2020naj}. If the 
elliptic flow buildup in the hadronic phase is additive with respect to the 
flow at the beginning of this phase, then an implementation of collective 
behavior at the partonic phase would only need to account for a fraction of 
the final-state elliptic flow. 

\begin{figure}[htbp!]
\includegraphics[width=7.8cm]{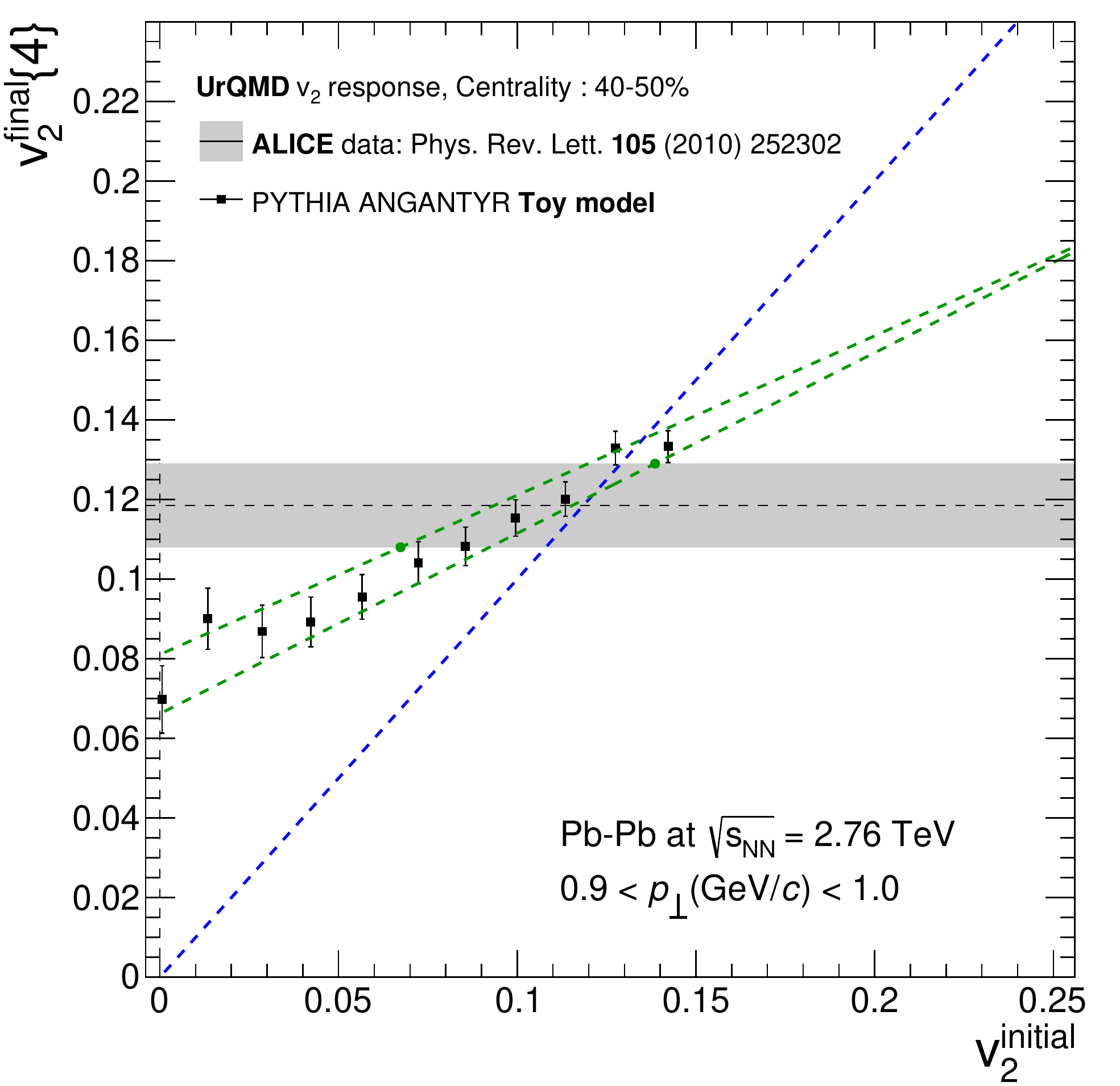}  
\includegraphics[width=7.8cm]{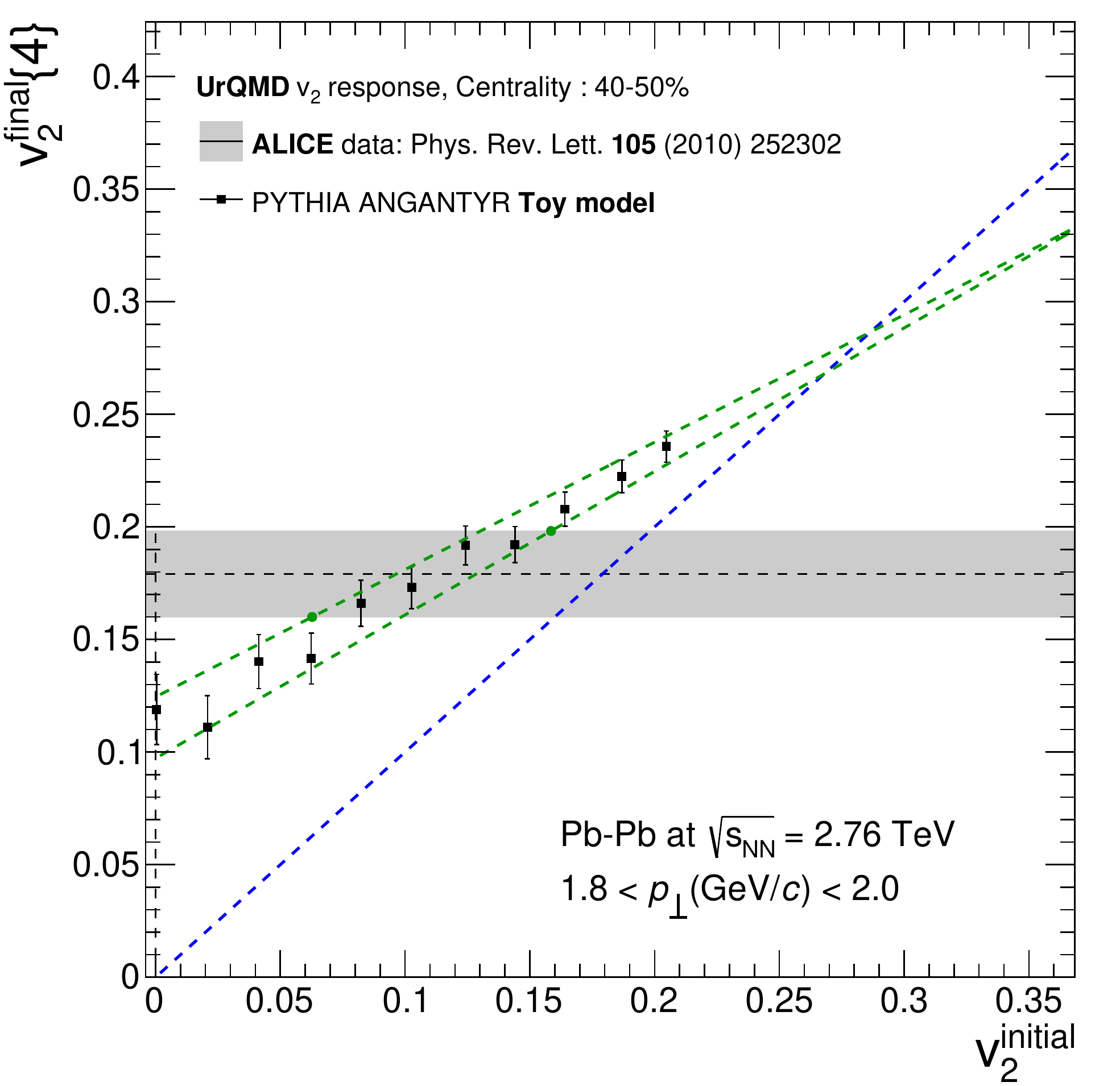}
\caption{Final-state $v_{2}\{4\}$ versus pre-hadronic-phase $v_{2}$ 
in the hydrodynamics-based and \pythia-based models for two selected \pt 
intervals. The black points correspond to setups in which a
pre-hadronic-phase $v_{2}^{\rm{Fit}}$ was added to \pytang  output by slightly 
rotating the hadrons' transverse momentum towards the event plane, as 
described in the text. The dashed green lines is a linear fit to the 
minimum and maximum value of $v_2^{\rm Final}\{4\}$ allowed by the 
error bars and are used to build Fig~.\ref{fig:minmax_flow}.}
\label{fig:toymodel}
\end{figure}

\begin{figure}[htbp!]
  \includegraphics[width=7.8cm]{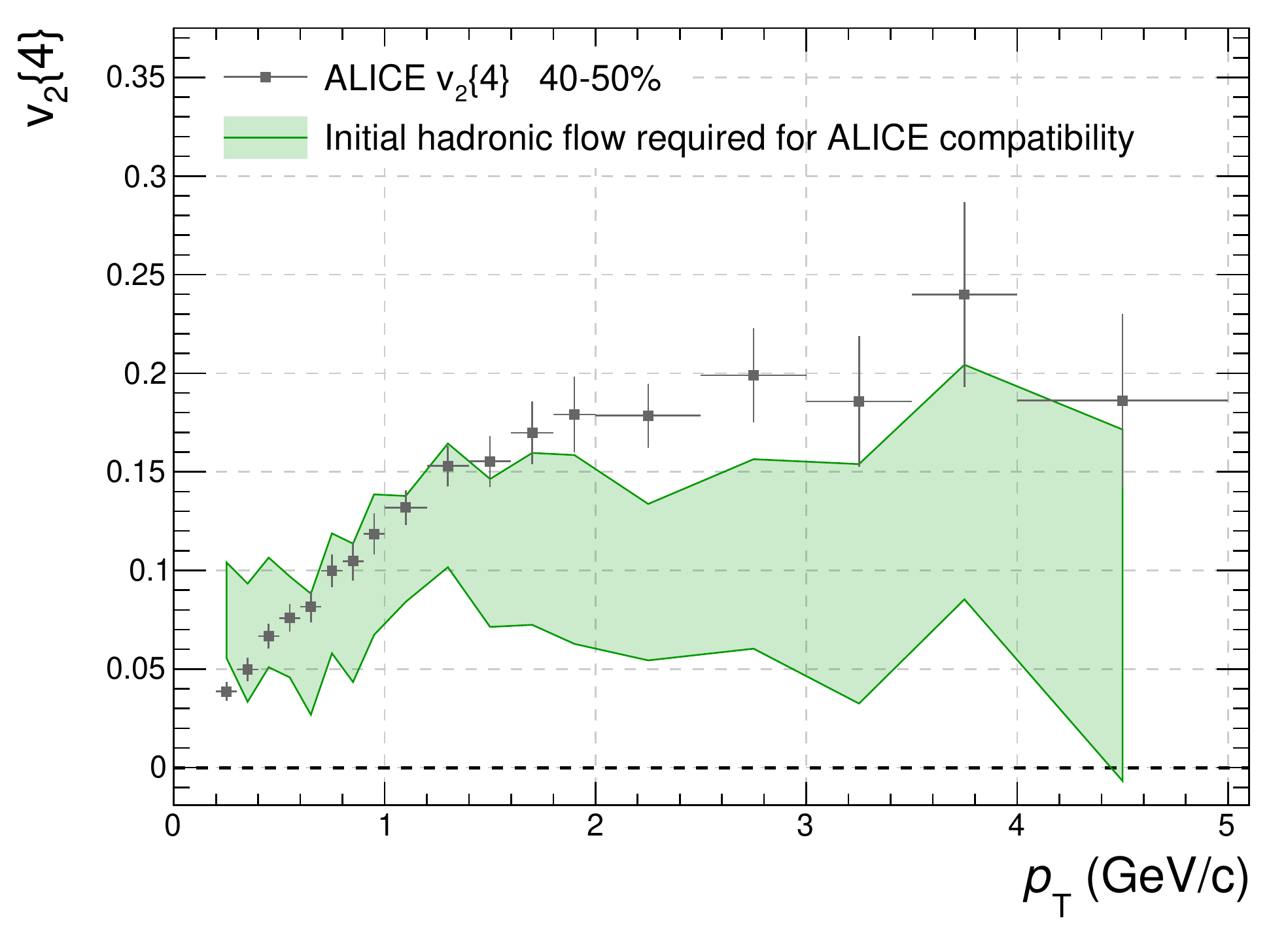}  
  \caption{Range of acceptable values that a partonic phase like \pytangur 
  should have as to reproduce experimental data (green band). We also show the 
  elliptic flow values measured experimentally as reference.}
  \label{fig:minmax_flow}
\end{figure}

In order to test \urqmd response, we developed a setup in which the transverse
momenta obtained from the \pytang generator are rotated before the \urqmd 
evolution in such a way that a configurable $v_2(p_\perp)$ could be set as desired. 
This artificial $v_2(p_\perp)$ is introduced following the relation
\begin{equation}
    \begin{split}
    v_2(p_\perp) & = \frac{A\pt}{1+Bp_\perp\left(1+Cp_\perp^2\right)}\,, \\
           B & = 0.315\, (\si{\giga \electronvolt \per \clight})^{-1}\,, \\
           C & = 0.112\, (\si{\giga \electronvolt \per \clight})^{-2}\,,
  \end{split}
  \label{eq:toy_model}
\end{equation}
where $A$ is a parameter that can be freely scaled to obtain various levels of initial flow. The values of $B$ and $C$ were obtained by fitting Eq.~\ref{eq:toy_model} to 
$v_2$ obtained from the \pytangur model in the centrality class 
40-50\%, which exhibits the same momentum dependence as the
measured elliptic flow in Pb-Pb collisions while allowing for 
a fit to be performed in narrow $p_{\perp}$ intervals. The scale parameter $A$ can be changed to increase or decrease the 
magnitude of the introduced elliptic flow. We simulated with 11 different values of $A$ and show 
how the final, post-hadronic-phase $v_{2}\{4\}$
changes as a function of the the  $v_{2}$ introduced at hadronization time.
The gray band shows us the value of $v_2\{4\}$ as computed by the ALICE 
collaboration in Ref.~\cite{Aamodt:2010pa}. Figure~\ref{fig:toymodel} indicates
that there is a dependence of \urqmd response with respect to the transverse momentum.
The rate with which the final-state elliptic flow increases seems to be 
smaller at low\pt than at high \pt. The result is that the pre-hadronic phase 
needs to generate almost the same amount of flow as in the experiment in the 
low-\pt region. But for higher transverse momentum, only 50\% of the final
state flow needs to be present. A complete study of the allowed range of initial
flow (in centrality 40-50\%) can be found in Figure~\ref{fig:minmax_flow}. This indicates that a model
like \pytang should generate flow at hadronization time with a weaker 
dependence of transverse momentum than what is observed experimentally, providing guidance to further \pytang 
developments such as string shoving \cite{Bierlich:2020naj}. 

\section{Conclusion}
In this work, we have discussed three basic aspects of \pytangur simulations: the nuclear modification
factor, two-particle correlations and elliptic flow. 
The fact that the nuclear modification factor calculated using \pytangur simulations follows the qualitative trend of the measured $R_{\rm{AA}}$ at intermediate $p_\perp$ once hadronic
interactions are considered is an intriguing observation in itself.
As explained in the introduction, values for $R_{\rm{AA}}$ below unity are generally taken as a clear indication of QGP formation, and models incorporating QGP formation are, as a general rule, needed to describe the data. In \pytangur there is no assumption of a QGP phase, but nevertheless several effects contribute to the description of $R_{\rm{AA}}$.

The high-$p_\perp$ part of the spectra generated by \pytang deviates from the simple binary
scaling which is normally expected \cite{Miller:2007ri}, as seen also in the fact that the $R_{\rm{AA}}$ is constant but below unity for high $p_{\perp}$ even without hadronic interactions. As previously explained, \pytang
makes a distinction between various types of nucleon--nucleon interactions, which will contribute 
differently to high-$p_{\perp}$ particle yields, and as a result the high-$p_{\perp}$ $R_{\rm{AA}}$ 
does not converge to unity even if hadronic interactions are disabled. This effect is responsible 
for the majority of the deviation from unity, as seen in \figref{fig:nucmod}.
There are, however, model uncertainties associated with this effect. While the treatment of secondary
absorptive sub-collisions similar to diffractive excitations can be theoretically and numerically motivated \cite{Bierlich:2018xfw},
it is, as mentioned earlier, not clear that it will exactly reproduce the phenomenology of an
interleaved shower plus color reconnection. 
An obvious next step would be to study the (absence of) nuclear modification
in p-Pb collisions within \pytangur. However, in that case, model 
uncertainties are much larger than in \AA collisions. It was shown in 
ref. \cite{Bierlich:2018xfw} that secondary collisions contribute between 25-40\% in Pb-Pb
collisions at $\sqrt{s_{\rm{NN}}}$ = 2.76~TeV, while for p-Pb collisions at $\sqrt{s_{\rm{NN}}}$ = 5.02~TeV they contribute between 50-85\%.

An equally important conclusion from this work is the influence from the hadronic rescattering phase
on $R_{\rm{AA}}$. Since strings have an average life-time $\langle \tau^2 \rangle \approx 2$ fm/$c$, the hadronic
phase is longer, and with a more dense initial condition, than for hydrodynamic simulations where a QGP phase lives
for up to an order of magnitude longer. We have shown that a hadronic rescattering phase with such an early starting 
time modifies $R_{\rm{AA}}$ up to a factor 3 for intermediate $p_\perp$ particles in central Pb-Pb events. 
From this model, we can interpret that at intermediate $p_{\perp}$, hadronic interactions lead to a 
suppression of particle yields as hadrons lose significant momentum to more abundant low-$p_{\perp}$ particles. 
However, this effect subsides for very large transverse momentum, such that the $R_{\rm{AA}}$ 
will eventually converge to the value without hadronic interactions. 
This is because, in the hadron vertex model, higher-$p_{\perp}$ values correlate with more displaced hadron creation
vertices due to the linear relationship between space-time and momentum in \eqref{eq:spacetime}, visible also in Fig.~\ref{fig:radiusdistrib}. 
Since higher-$p_{\perp}$ particles are then increasingly displaced, these 
are less likely to interact with the remainder of the system. While this effect is smaller in magnitude 
than the deviation from binary scaling, it is crucial to recover the minimum of the nuclear modification 
factor at around 5~GeV/$c$ and the subsequent rise at high momenta.

It has to be noted that the intermediate to high-$p_{\perp}$ suppression studied in this paper is fundamentally different than 
the one that would result from models such as JETSCAPE or JEWEL. In the latter, jet quenching is a 
phenomenon associated to partonic energy loss, which would lead to the suppression of an entire
high-momentum jet, while in the former, individual hadrons
lose momentum after hadronization. Experimentally, these 
two scenarios can be distinguished using techniques 
such as two-particle correlations and dijet asymmetry measurements. It is with this motivation that
we pursued the studies shown in Fig.~\ref{fig:2pc}, which indicate that the high-momentum particle suppression
from hadronic rescattering follows, in fact, again the same qualitative trend as what was observed by the STAR collaboration at 
RHIC \cite{Adler_2003,PhysRevLett.91.072304}, with the near-side jet being mostly unaltered and the away-side jet being suppressed due to a larger number of 
interactions with the hadronic medium. These findings are complementary to recent studies
on the effect of hadronic rescattering on jet shapes \cite{Dorau_2020} which also indicate that the 
hadronic phase does have significant impact on high-$p_\perp$ physics observables. 

To further elaborate on the loss of dijet structures at high $p_\perp$, we also calculated the $p_\perp$-differential $v_{2}\{2\}$
with and without interactions and compared the results to measurements by the CMS collaboration. While the predictions without 
rescattering overshoot the CMS measurements very significantly, the addition of hadronic interactions reduces the $v_{2}\{2\}$ 
very significantly, bringing it closer to CMS data even if the overall magnitude of the $v_{2}\{2\}$ is underpredicted. 
Furthermore, we studied \urqmd response by artificially 
introducing flow at the beginning of the hadronic phase
and have shown that, while flow must still be generated 
during the partonic phase, its overall momentum 
dependence may be milder as the hadronic phase still contributes
to the final flow. This observation sets the stage for future
developments of the \pytang model, notably the inclusion 
of string shoving~\cite{Bierlich:2016vgw, Lonnblad:2019dyj}. 
A more comprehensive 
study of flow coefficients from the \pytangur model suggests itself as the next step of this work, but lies beyond the scope 
of this particular manuscript. Another topic of interest that may be pursued in the future is
the evolution of identified particle ratios with centrality, in which case inelastic scatterings
in the hadronic phase may have a significant impact in high-multiplicity heavy-ion collisions. 

We also take this opportunity to highlight further work done to include hadronic rescattering in
pp collisions even within the PYTHIA generator itself \cite{sjstr2020framework}. A natural extension
of these studies would be to use the same machinery in Pb-Pb collisions, which might provide several
advantages over \urqmd - for instance, with the possibility of treating also charmed hadrons 
during the hadronic phase. 

The findings of this paper prompt present and future experimental studies to
put more emphasis in the construction of baseline, 
no-QGP models, as a certain fraction of the observations normally exclusively 
associated with the QGP may have its origin elsewhere. Phrased differently, this work conclusively demonstrates
that comparing experimental data with an incoherent superposition of proton-proton collisions
is not a valid exercise to fully isolate QGP-specific equilibration signatures.

\section{Acknowledgements}

We thank Marcus Bleicher for insightful discussions.

The authors would further like to acknowledge support from individual funding bodies:
CB was supported by Swedish Research Council, contract number 2017-003.
AVS, DDC, WMS and JT were supported by FAPESP project number 17/05685-2 and 2009/54213-0. 
AVS would also like to acknowledge CAPES for the support under Finance Code 001.
This research used the computing resources and assistance of the John David
Rogers Computing Center (CCJDR) in the Institute of Physics "Gleb
Wataghin", University of Campinas.

\bibliographystyle{ieeetr}
\bibliography{paper}
\end{document}